\documentclass[10pt,conference]{IEEEtran}
\usepackage{cite}
\usepackage{amsmath,amssymb,amsfonts}
\usepackage{algorithmic}
\usepackage{graphicx}
\usepackage{textcomp}
\usepackage[table]{xcolor}
\usepackage[hyphens]{url}
\usepackage{fancyhdr}
\usepackage[hidelinks]{hyperref}
\usepackage{multirow}
\usepackage{tablefootnote}

\usepackage{makecell}
\usepackage{enumitem}

\usepackage[english]{babel}

\definecolor{rowgray}{gray}{0.95}
\newcommand{\hpcayear}{2026}

\newcommand{\hpcasubmissionnumber}{1567}

\title{zkPHIRE: A Programmable Accelerator for \\ \underline{Z}\underline{K}\underline{P}s over \underline{HI}gh-deg\underline{R}ee, \underline{E}xpressive Gates}
\def\hpcacameraready{} % Uncomment to build camera-ready version

\newcommand\hpcaauthors{Alhad Daftardar$^*\dagger$, Jianqiao Mo$^*\dagger$, Joey Ah-kiow$\dagger$, Benedikt B\"unz$\ddagger$, Siddharth Garg$\dagger$, Brandon Reagen$\dagger$}
\newcommand\hpcaaffiliation{New York University Tandon School of Engineering$\dagger$\\New York University Courant Institute of Mathematical Sciences$\ddagger$}
\newcommand\hpcaemail{\{ajd9396, jm8782, ja4844, bb, sg175, bjr5\}@nyu.edu \\
$^*${These authors contributed equally to this work}
}

\begin{document}

%%%%%%%%%%%%%%%%%%%%%%%%%%%%%%%%%%%%%
%%%%%%%%%% -- DO NOT MODIFY -- %%%%%%%%%%
%%%%%%%%%%%%%%%%%%%%%%%%%%%%%%%%%%%%%

\author{
  \ifdefined\hpcacameraready
    \IEEEauthorblockN{\hpcaauthors{}}
      \IEEEauthorblockA{
        \hpcaaffiliation{} \\
        \hpcaemail{}
      }
  \else
    \IEEEauthorblockN{\normalsize{HPCA \hpcayear{} Submission
      \textbf{\#\hpcasubmissionnumber{}}} \\
      \IEEEauthorblockA{
        Confidential Draft \\
        Do NOT Distribute!!
      }
    }
  \fi 
}

% Heading and footer for title page
\fancypagestyle{camerareadyfirstpage}{%
  \fancyhead{}
  \renewcommand{\headrulewidth}{0pt}
  \fancyhead[C]{
    \ifdefined\aeopen
    \parbox[][12mm][t]{13.5cm}{\hpcayear{} IEEE International Symposium on High-Performance Computer Architecture (HPCA)}    
    \else
      \ifdefined\aereviewed
      \parbox[][12mm][t]{13.5cm}{\hpcayear{} IEEE International Symposium on High-Performance Computer Architecture (HPCA)}
      \else
      \ifdefined\aereproduced
      \parbox[][12mm][t]{13.5cm}{\hpcayear{} IEEE International Symposium on High-Performance Computer Architecture (HPCA)}
      \else
      \parbox[][0mm][t]{13.5cm}{\hpcayear{} IEEE International Symposium on High-Performance Computer Architecture (HPCA)}
    \fi 
    \fi 
    \fi 
    \ifdefined\aeopen 
      \includegraphics[width=12mm,height=12mm]{ae-badges/open-research-objects.pdf}
    \fi 
    \ifdefined\aereviewed
      \includegraphics[width=12mm,height=12mm]{ae-badges/research-objects-reviewed.pdf}
    \fi 
    \ifdefined\aereproduced
      \includegraphics[width=12mm,height=12mm]{ae-badges/results-reproduced.pdf}
    \fi
  }
  %\fancyfoot[L]{\hpcapubid{} \copyright \hpcayear{} IEEE}
  \fancyfoot[C]{}
}
% Heading and footer for remaining pages
\fancyhead{}
\renewcommand{\headrulewidth}{0pt}
%\fancyhead[C]{\hpcayear{} IEEE International Symposium on
% High-Performance Computer Architecture (HPCA)}

\maketitle

%Enables the camera ready header and footer
\ifdefined\hpcacameraready 
  \thispagestyle{camerareadyfirstpage}
  \pagestyle{empty}
\else
  \thispagestyle{plain}
  \pagestyle{plain}
\fi

\newcommand{\hpcaheight}{0mm}
\ifdefined\eaopen
\renewcommand{\hpcaheight}{12mm}
\fi
  
\begin{abstract}
Zero-Knowledge Proofs (ZKPs) have emerged as a powerful tool for secure and privacy-preserving computation. ZKPs enable one party to convince another of a statement’s validity without revealing anything else. This capability has profound implications in many domains, including machine learning, blockchain, image authentication, and electronic voting. Despite their potential, ZKPs have seen limited deployment because of their exceptionally high computational overhead, which manifests primarily during proof generation. To mitigate these overheads, a (growing) body of researchers has proposed hardware accelerators and GPU implementations of both kernels and complete protocols. Prior art spans a wide variety of ZKP schemes that vary significantly in computational overhead, proof size, verifier cost, protocol setup, and trust. The latest and widely used ZKP protocols are intentionally designed to balance these trade-offs.
One particular challenge in modern ZKP systems is supporting complex, high-degree gates using the SumCheck protocol. We address this challenge with a novel programmable accelerator to efficiently handle arbitrary custom gates via SumCheck. Our accelerator achieves upwards of $1000\times$ geomean speedup over CPU-based SumChecks across a range of gate types. We include this unit in zkPHIRE, a programmable, full-system accelerator that accelerates the HyperPlonk protocol. zkPHIRE achieves $1486\times$ geomean speedup over CPU and $11.87\times$ geomean speedup over the state-of-the-art at iso-area. 
Together, these results demonstrate compelling performance while scaling to large problem sizes (upwards of $2^{30}$ constraints) and maintaining small proof sizes ($4-5$ KB).

\end{abstract}

\maketitle

\section{Introduction}
\label{sec:Introduction}
Zero-Knowledge Proofs (ZKPs) are cryptographic protocols that allow a prover to convince one or more verifiers that a claim is correct without revealing any information beyond the validity of the claim. For example, a prover can demonstrate possession of a valid authentication key without disclosing the key itself. 
ZKPs have many compelling applications, including privacy-preserving authentication, trustworthy AI, blockchain scalability, and recently, private age verification \cite{google_zkp}.
A major limitation of state-of-the-art ZKP protocols is their high computational cost--generating a single proof on a CPU can take minutes to hours. As a result, there is growing interest in accelerating ZKPs, both with programmable solutions (e.g., GPUs~\cite{gzkp, distmsm}), and ASICs~\cite{pipezk, szkp, rezk, nocap, zkspeed2025}.

ZKP protocols continue to evolve and improve.
Many have been proposed, offering different trade-offs in runtime, proof size, verifier time, setup (universal vs. application-specific), and security guarantees.
A major family of ZKP protocols relies on the SumCheck protocol\cite{sumcheck}. 
SumCheck-based protocols have $O(N)$ prover time, as compared to NTT-based protocols (a popular alternative\cite{groth16,gabizon2019plonk}), which require $O(N \log(N))$ prover complexity.
The SumCheck protocol is highly flexible and is at the core of many classic and modern ZKP protocols~\cite{goldwasser2015delegating,spartan,wahby2018doubly,hyperplonk,orion,whir,basefold}, which have relatively fast prover time.
 While SumChecks were part of previously accelerated systems~\cite{verifiable_asics,nocap,zkspeed2025}, they studied specific instantiations of the SumCheck protocol rather than the protocol family more generally. 
Another major distinguishing factor between protocols is the type of polynomial commitment scheme\cite{kzg_pcs} that is used. 
 Some of these protocols, such as  
\textbf{Spartan}~\cite{spartan}\textbf{+Orion}~\cite{orion}, use more lightweight Merkle tree commitments at the cost of large proofs (many MBs), while
\textbf{HyperPlonk}~\cite{hyperplonk} uses a more computationally demanding commitment scheme that produces
significantly smaller proofs (a few kBs). 

ZKP protocols typically encode computations as arithmetic circuits (ACs) -- directed acyclic graphs where each node (or gate) enforces a local algebraic constraint, such as addition ($f = a+b$) or multiplication ($f = a\cdot b$) over a finite field. We generally denote the number of gates by $N$. These circuits are then arithmetized by translating gates into polynomial constraints, which are later verified using polynomial checks for correctness. The SumCheck protocol is used to reduce checking $N$ gates to checking a single, randomized gate formula.
A key feature of newer SumCheck-based protocols is the use of \textit{custom, high-degree gates} that encapsulate more computation per gate by using higher-degree polynomial constraints (e.g., $f = a\cdot b^5 + c^2$).
This reduces the overall number of gates needed to represent a computation, resulting in large efficiency gains. The move to custom gates is exemplified in the Halo2 arithmetization language\cite{halo2}. Halo2 enables a circuit designer to specify the types of gates they want to use \emph{per circuit}. 

However, this level of customizability introduces significant hardware challenges. Unlike NTTs, which have fixed, regular dataflows, SumCheck computations vary with each gate’s structure and polynomial degree. High-degree polynomials require more modular multiplications and reductions, leading to increased bandwidth pressure, varied memory access patterns, and more complex scheduling. Designing efficient SumCheck hardware thus involves a trade-off: either build fixed-function units for specific gates, or develop a general, programmable architecture that can dynamically adapt to diverse polynomials without significantly compromising performance. 

zkSpeed~\cite{zkspeed2025} exemplifies the first approach, using dedicated hardware for Vanilla gates (additions and multiplications) within the HyperPlonk protocol. While this provides good performance for a fixed gate set, it limits flexibility: zkSpeed’s SumCheck unit cannot easily handle complex custom gates like those seen in the Halo2 arithmetization, reducing its applicability across general ZKP workloads. Furthermore, zkSpeed's reliance on large on-chip scratchpads limits scalability, as its memory footprint grows with the number of gates, becoming impractical beyond $2^{24}$ gates, a scale increasingly common in emerging ZKP applications.

Generalizable SumCheck cores are needed beyond ZKPs, too. In verifiable computing, SumCheck serves as the core verification mechanism in protocols that provide integrity guarantees without requiring zero knowledge. However, previous work in this area, including verifiable ASICs~\cite{verifiable_asics}, has also relied on custom SumCheck units with fixed function, again restricting applicability and reuse.

In this work, we adopt the second approach with \textit{\textbf{zkPHIRE}}, introducing a programmable SumCheck accelerator that supports arbitrary polynomial structures and gate types. zkPHIRE addresses the irregularity and data reuse challenges intrinsic to SumCheck with fused compute pipelines, tree-based interconnects for efficient reductions, and flexible scheduling to handle diverse polynomial structures and degrees. This flexibility opens the door to a wide range of applications across ZKP and verifiable computing. We evaluate our programmable SumCheck unit on a diverse array of gate types, and then embed it into a full-system accelerator, zkPHIRE. zkPHIRE improves upon zkSpeed with more efficient memory organization and module abstraction that allows hardware to be reused across protocol steps. We conduct a protocol-level evaluation using the HyperPlonk with custom gates as a representative case study.
Our key contributions are:
\begin{itemize}
    \item A novel, programmable SumCheck unit that supports arbitrary high-degree multilinear polynomials, enabling compatibility with custom gate types.
    \item The \textit{first} hardware evaluation of high-degree gates that offer significant performance and scalability benefits.
    \item A unified Multifunction Forest and optimized modular inverse unit that (respectively) save 15.2\% and 4.2$\times$ area with improved utilization compared to prior work~\cite{zkspeed2025}.
    \item A full-system accelerator, zkPHIRE, that achieves $11.87\times$ geomean speedup over zkSpeed, $1486\times$ geomean speedup over CPU, and supports succinct proofs for problem sizes up to $2^{30}$ nominal constraints.

\end{itemize}

\section{Background}
\label{sec:Background}

\subsection{Zero-Knowledge Proofs}
Today, the state-of-the-art ZKP protocols are zero-knowledge Succinct Non-interactive Arguments of Knowledge (zkSNARKs). They have three key properties:
\begin{itemize}
    \item {\textbf{zero-knowledge:}} The verifier learns nothing beyond the validity of the statement.
    \item {\textbf{succinctness:}} The proof is short and verified quickly.
    \item {\textbf{non-interactivity:}} The proof is a single message, eliminating the need for back-and-forth communication.
\end{itemize}

Modern zkSNARKs combine an Interactive Oracle Proof (IOP) and a Polynomial Commitment Scheme (PCS). The IOP encodes the witness and computation as polynomials, which the verifier checks via probabilistic queries. Applying the Fiat-Shamir heuristic~\cite{fiat_shamir} makes public-coin IOPs non-interactive. The PCS lets the prover commit to a polynomial and later open it at chosen points, ensuring both binding (the prover cannot change the committed polynomial) and hiding (nothing about the polynomial is revealed until it is opened).

Different zkSNARKs balance proving time, proof size, and security assumptions based on their IOP and PCS designs. Groth16~\cite{groth16} offers fast verification and tiny proofs, ideal for blockchains, but requires a circuit-specific (i.e., per-application) trusted setup and scales poorly with circuit size. Code-based protocols~\cite{orion} use error-correcting codes and Merkle commitments to achieve transparency and post-quantum security, but at the cost of multi-megabyte proofs and slower verification. HyperPlonk~\cite{hyperplonk} trades off slightly larger proofs for universal setup and linear-time proving.

\subsection{Multi-Scalar Multiplication}
Multi-Scalar Multiplication (MSM) is a key kernel used in \textit{polynomial commitments} and relies on elliptic curve cryptography.
MSMs involve computing a linear combination (i.e., dot product) of scalar and (2D or 3D) point pairs on an elliptic curve.
Given scalars \( k_1, k_2, \ldots, k_n \) and corresponding elliptic curve points \( P_1, P_2, \ldots, P_n \), the goal is to efficiently compute $s = \sum_i k_iP_i$.
MSMs are computationally intensive because each term \( k_iP_i \) involves a point-scalar multiplication -- a costly elliptic curve operation. 
MSMs constitute a significant portion of the total runtime in many zkSNARKs, including Groth16 and HyperPlonk. 
As a result, prior work has focused on accelerating MSMs using both application-specific and programmable platforms~\cite{szkp, pipezk, cuzk, gzkp, distmsm, priormsm, rezk}.
Rather than computing each scalar multiplication directly, optimized approaches such as Pippenger’s algorithm~\cite{pippenger} are commonly employed.
Pippenger’s algorithm improves efficiency by restructuring the computation to rely more heavily on faster operations like point addition and point doubling.

\subsection{SumCheck}
\label{sec:sumcheck_intro}
SumCheck~\cite{sumcheck} is a round-based interactive protocol where a prover $\mathcal{P}$ convinces a verifier $\mathcal{V}$ that the sum of a multivariate polynomial, $f(X_1, \ldots,X_\mu)$, over the \textit{boolean hypercube} (all boolean values of $X_i \ldots X_\mu$) equals a claimed value $C$.
The protocol proceeds in $\mu$ rounds, one per variable. 
In round $i$, $\mathcal{P}$ sends a univariate polynomial
\begin{equation*}
    s_i(X_i) = \sum_{x_{i+1} \in \{0,1\}} \cdots \sum_{X_\mu \in \{0,1\}} 
    f(r_1, \ldots, r_{i-1}, X_i, \ldots, X_\mu)
\end{equation*}
where $r_1, \ldots, r_{i-1}$ are $\mathcal{V}$’s random challenges from prior rounds. $\mathcal{V}$ checks that $s_i(0) + s_i(1)$ equals the previous round's claim (or the initial claim $C$ for round $1$).
$\mathcal{V}$ then chooses a random field element $r_i$ and sends it to $\mathcal{P}$, who sets $X_i = r_i$ for the next round.
After $\mu$ rounds, $\mathcal{V}$ evaluates $f$ at $(r_1, \ldots, r_\mu)$ and accepts if all checks pass.

ZKPs execute SumCheck over \textit{multilinear extensions} (MLEs) -- polynomials with degree at most $1$ in each variable. MLEs represent multivariate functions compactly and, in hardware, can be stored as flat lookup tables indexed by binary inputs. Throughout this paper, we use the terms ``MLEs," ``MLE tables," and ``polynomials" interchangeably.

\subsubsection{\textbf{Polynomials and Gates}}
In ZKPs, polynomials encode and help enforce the correctness of the computation being proven. A popular encoding scheme is Plonk \cite{gabizon2019plonk}. Plonk's \textit{arithmetization} scheme converts the underlying computation into an arithmetic circuit consisting of gates. Plonk circuits support additions, multiplications, conditionality, and equality checks. They can be represented by the composite polynomial 
\begin{equation*}
\label{eq:plonk}
    f_{plonk} = q_Lw_1 + q_Rw_2 + q_Mw_1w_2 - q_Ow_3 + q_C
\end{equation*}
The composite polynomial $f_{plonk}$ has 5 terms and a total of 8 unique, constituent polynomials.
These constituent polynomials are the selectors (or enables) $q_i$ and witnesses $w_i$, and they are all multilinear.
Notably, 
% $f$
$f_{plonk}$
is not itself multilinear; rather, it is composed of multilinear polynomials.

Each of these polynomials encodes a function of a multivariate input vector $X=[X_1, X_2, \ldots, X_\mu]$, which encodes the gate index in binary. For example, if $\mu = 3$, there are 8 gates, and gate index 3 would be represented by evaluating $f_{plonk}(1, 1, 0)$. For a multiply operation, we would set $q_M(1, 1, 0) = 1$, $q_O(1, 1, 0) = 1$ and the other enable polynomials to $0$. Then, gate 3 would execute correctly if and only if $f_{plonk}(1, 1, 0) = 0$, that is, 

\begin{equation*}
w_1(1,1,0)\cdot w_2(1,1,0) - w_3(1,1,0) = 0
\end{equation*}
Representing circuits as polynomials lends itself naturally to SumCheck-based provers where we can check (with additional constraints) that each gate $f_{plonk}(X)$ evaluates to $0$ and that
\begin{equation*}
\sum_X f_{plonk}(X) = 0 
\end{equation*}

\subsubsection{\textbf{High-degree, Expressive Gates}}
The challenge of using Plonk gates is that all computations must be mapped to only what is supported by the Plonk polynomial, 
similar to mapping program code to a RISC ISA.
In some cases, this results in circuits that require millions ($>2^{20}$) of gates, and in turn, polynomials with millions of evaluations.
Recent protocols and libraries ~\cite{halo2, hyperplonk} support \textit{expressive, high-degree} gates, which encode complex operations and significantly reduce gate count.
For example, HyperPlonk’s \textit{Jellyfish} gate captures common patterns such as hashing; across benchmarks, it reduces gate counts by up to 32× compared to less expressive \textit{Vanilla} Plonk gates.
Similarly, Halo2 uses custom gates for elliptic curve operations. However, these expressive gates increase the complexity of the SumCheck algorithm used to verify them.

\subsubsection{\textbf{SumCheck on Multilinear Polynomial Products}}
SumCheck over a single multilinear polynomial is relatively straightforward.
Consider a polynomial $f(X_1, X_2, X_3)$ with $\mu = 3$ and its MLE table representation.
In the first round, the prover computes the univariate polynomial $s_1(X_1)$ by summing entries with $X_1 = 0$ and $X_1 = 1$, yielding two evaluations $s_1(0)$ and $s_1(1)$ sent to the verifier.
The prover then computes an \textit{MLE Update} by fixing $X_1$ to a random challenge $r_1$, typically generated by hashing the round evaluations (e.g., with SHA3).
For each $(X_2, X_3)$, the entries $f(0, X_2, X_3)$ and $f(1, X_2, X_3)$ define a line
in $X_1$, which is extrapolated to $r_1$: $f(r_1,X_2, X_3) = f(0, X_2, X_3)\cdot(1-r_1) + f(1, X_2, X_3)\cdot r_1$.
This halves each MLE table before the next SumCheck round.

In recent SumCheck-based ZKP protocols \cite{spartan, orion, hyperplonk, jolt}, we typically work with multi-term compositions of multilinear polynomials, like $f_{plonk}$ in \autoref{eq:plonk}. In these protocols, we are given \textit{only} the constituent polynomials and their composition structure, and then must perform SumCheck over the composition. These SumChecks are naturally more complicated.
In general, to run SumCheck on a $d$-degree polynomial, we require $d+1$ evaluations for $s_i(X_i)$ after each $i^{th}$ round of SumCheck.
For example, $f_{plonk}$'s overall degree is $3$, so \textit{$4$} evaluations are needed per SumCheck round.
These evaluations must be computed for each individual MLE before computing term-wise products and summing across terms.

\begin{figure}[t!]
\centerline{\includegraphics[width=1.02\columnwidth]{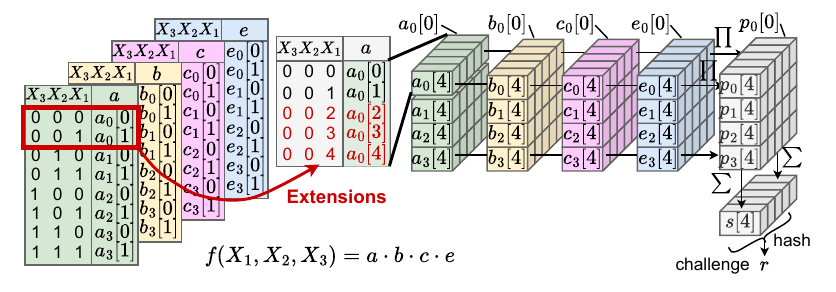}}
\caption{The SumCheck dataflow of size-$2^{3}$ polynomial $f(X)=abce$.
For a degree-4 polynomial $f$, every MLE evaluation pair is extended to five evaluations.
Extended evaluations are multiplied, summed, and hashed to get a challenge that is used to update MLE tables for the next SumCheck round. }
\label{fig:SumCheck_dataflow_example}
\end{figure}

\autoref{fig:SumCheck_dataflow_example} shows the dataflow for one polynomial product term.
Each pair of entries in an MLE table actually represents the evaluations at $X_i = 0$ and $1$; these values are then \textit{extended} to $X_i = 2, \ldots, d$, using the same formula as MLE Update (hence multilinear \textit{extension}).
Then, the extensions for all polynomials in a term are multiplied together.
These products are summed down the MLE table, yielding $d+1$ values for a term.
When there is more than one term, 
this process will be repeated for each term, with the $d+1$ evaluations summed across terms to form the final polynomial $s_i(X_i)$ for round $i$.

We now present our generalizable SumCheck accelerator.

\section{The Programmable SumCheck Accelerator}

\subsection{Challenges}

At its core, SumCheck offers several degrees of parallelism. 
Different MLE tables can compute their extensions in parallel, extensions for different MLE pairs \textit{within} a table can be computed in parallel, and the products \textit{across} different extensions can be computed in parallel. Given a SumCheck polynomial, an optimal circuit can be constructed to maximize reuse and throughput, as done in zkSpeed \cite{zkspeed2025}.

In contrast, a generalizable SumCheck unit must support a diverse set of composite polynomial structures used to express different types of constraints or implement different kinds of operations. 
The unit must support an arbitrary number of terms and an arbitrary degree. This means that the optimal resource allocation (e.g., number of modular multipliers) for one composite polynomial may be suboptimal or wasteful for another one. Additionally, many composite polynomials have individual polynomials that repeat. For example, in $f = a\cdot b \cdot e + c \cdot e + e \cdot g $, the $e$ polynomial appears three times.
While we could parallelize all computations for the first term ($a \cdot b \cdot e$) before proceeding to the next term, this would leave out a clear opportunity for data reuse. Indeed, zkSpeed's design captures immediate reuse, but is tailored to HyperPlonk's Vanilla gate polynomials. zkSpeed overprovisions resources for simple polynomials, and cannot support complex composite polynomials.

\subsection{Datapath}
Our programmable SumCheck unit seeks to balance the trade-off between resource allocation and capturing immediate data reuse.
\autoref{fig:new_sumcheck}
shows the datapath 
for our programmable SumCheck unit. 
It consists of several processing elements (PEs), with MLE Update units, Extension Engines (EEs) and Product Lanes (PLs).
Instead of streaming all individual polynomials' MLE entries simultaneously, we proceed one term at a time, fetching a \textit{tile} of data from each MLE table and storing in local scratchpad buffers.
We allocate 16 scratchpad buffers, more than sufficient to accommodate polynomial structures we see in current ZKP systems.
The scratchpad buffers are banked to match the number of PEs for parallel accesses.
This ensures that individual polynomials reused in subsequent terms need not be fetched from off-chip memory for a given tile.

For a given term, the polynomials are read into the Update units and EEs, where their extensions are generated by a series of modular adders and subtractors.
Specifically, in round 1 of SumCheck, each PE reads two values at a time from MLE tables and directly feeds them to the EEs to generate extensions.
In all subsequent rounds, four values are read per MLE, because we pipeline the MLE updates directly into the extension computation.
Updating with four inputs ensures that two values are generated and fed to the EEs to match the 
same
throughput of round 1.
These updated values are also written to FIFOs that buffer writebacks to off-chip, forming the (halved) tables for the next SumCheck round.
In later rounds of SumCheck where MLE tables can fit fully on-chip, writebacks to off-chip are eliminated and updated values are directly routed to the scratchpad banks.

After each MLE's extensions are generated, they are packed into their respective groups, e.g., the extensions for all MLEs at $X_i = 2$ are routed together.
Each group of extensions is then fed into a Product Lane (PL).
Each PL is equipped with $E - 1$ fully pipelined modular multipliers, where $E$ is the number of EEs.
Each PL's output is accumulated into a register corresponding to the extension, e.g., all products at $X_i = 2$ are routed to register $2$.
We allocate 32 registers (up to degree 31); higher degrees are supported by storing in scratchpads.

\subsection{Mapping Polynomials to EEs}
\begin{figure}[t!]
\centerline{\includegraphics[width=0.98\columnwidth]{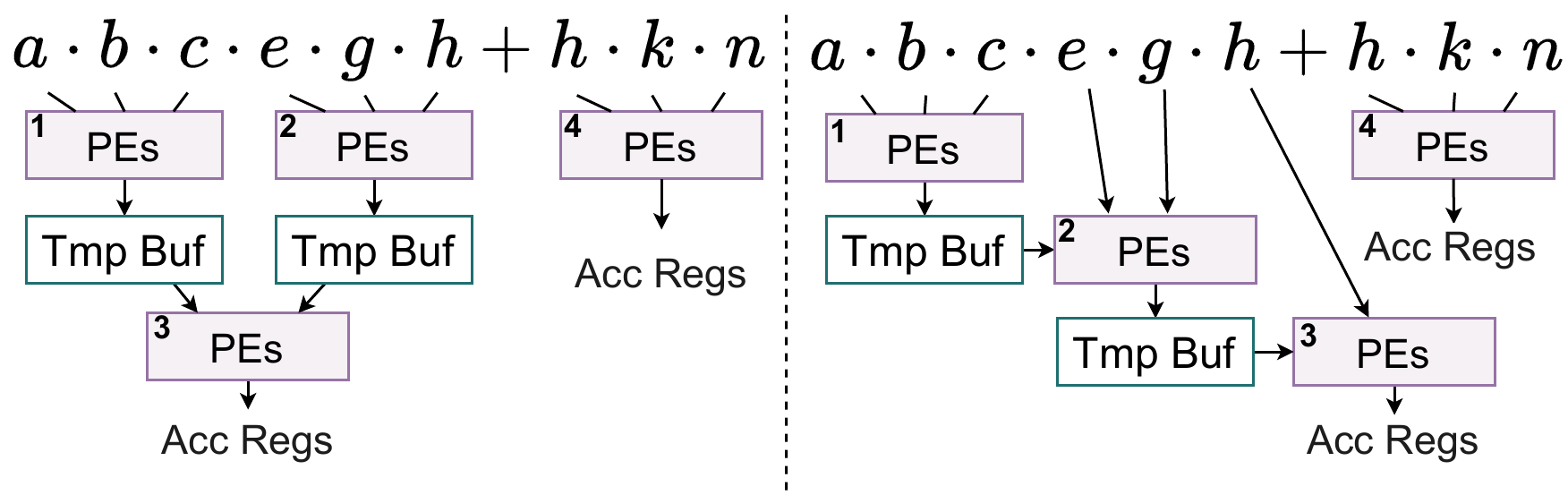}}
\caption{Graph decomposition used to schedule computation on our SumCheck unit. The method on the right minimizes the number of temporary MLE buffers. Numbers inside each node indicate the step in the schedule. When processing term $hkn$, $h$ is already prefetched during the prior step.}
\label{fig:scheduler}
\end{figure}

\begin{figure*}[t!]
    \centering
    % First figure in the left column
    \begin{minipage}{0.69\textwidth}
        \centering
        \includegraphics[width=\textwidth]{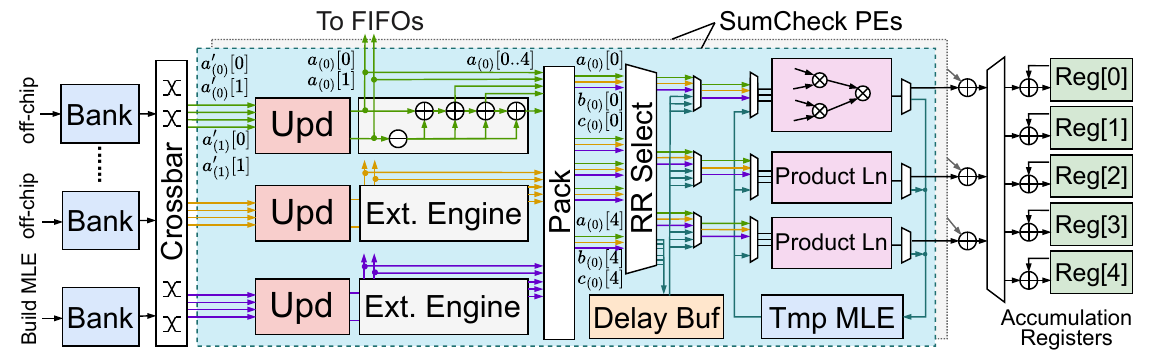}
        \caption{Programmable SumCheck Architecture.
        The Update unit updates the MLE $a^{\prime}$ to halved-size MLE $a$ for the next SumCheck round.
        For deg-4 term $abce$, Extension Engines generate five evaluations $0$-$4$.}
        \label{fig:new_sumcheck}
    \end{minipage}
    \hfill
    % Second figure in the right column
    \begin{minipage}{0.3\textwidth}
        \centering
        \includegraphics[width=\textwidth]{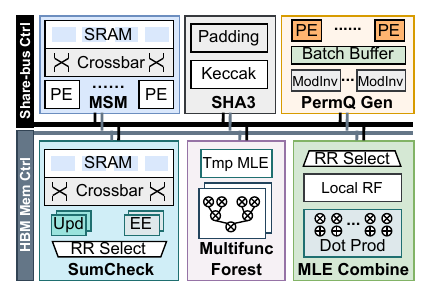}
        \caption{Overall Architecture of zkPHIRE.}
        \label{fig:overall_arch}
    \end{minipage}
\end{figure*}

We devise a scheduler that determines which MLE tables are fed to EEs and PLs using a graph-based decomposition (see \autoref{fig:scheduler}). 
In this example, we assume there are 3 EEs, so only 3 MLEs can be handled by a SumCheck PE at a time. 
The polynomial is adding a degree-6 term and a degree-3 term.
Since there are more MLEs than EEs, we must store the intermediate products for \textit{all} extensions, not just the initial extensions at $X_i = 0, 1$ that are provided from the original MLE tables. This is done using a dedicated \textit{Tmp} MLE buffer.

A natural way to schedule the MLEs is using the balanced tree-based approach on the left in \autoref{fig:scheduler}.
However, this would require storing two distinct intermediate products that would need to be combined.
As the polynomial degree grows, this tree-based approach would require more and more intermediate buffers.
Instead, we can use the accumulation-based schedule on the right.
This scheme only requires 1 Tmp MLE buffer that accumulates extension products within the same term.
Notice that this scheme uses the same number of steps as the balanced tree while minimizing temporary storage.

This way, the scheduler traverses the polynomial term-by-term, deciding which MLEs are handled by the PEs in each step.
In each step $j$, the PEs read the MLEs assigned to step $j$, while
a tile of data for each MLE needed in step $j + 1$ is prefetched if it is not already brought on-chip.
For example, the $h$ MLE is prefetched during step $2$ 
(on the right), 
while only the $k$ and $n$ MLEs are prefetched during step $3$.
For higher-degree polynomials, this scheme balances prefetch bandwidth across steps rather than front-loading it in the initial steps, as would occur with a binary-tree schedule.

\subsection{Mapping Extensions to PLs}
Our SumCheck unit must handle any number of extensions, and we may run into scenarios where there are $K$ extensions but only $P < K$ lanes. In this case, we have an initiation interval $II = K/P$ and use delay buffers to queue extra extensions. 
In \autoref{fig:new_sumcheck}, with $K=5$ and $P=3$, the first cycle maps extensions $0$–$2$ to PLs, while extensions $3$–$4$ are buffered for the next cycle.
Scheduling proceeds in a pipelined fashion, with buffered and incoming extensions interleaved across PLs.

\subsection{Programming the SumCheck unit}
The SumCheck unit is programmed via an automated scheduler given the target polynomial and hardware configuration.
Using the approach shown in Figure~2, the scheduler generates a list of computational steps, including MLE-to-EE mappings, prefetch ordering, and schedules for specific $(K, P)$ settings. 
These steps are annotated with signals for control registers (e.g., MLE bank selection, arbitration, bypassing update), address offsets, and FSM configuration. 
They are then loaded into on-chip controllers as instructions to guide SumCheck execution.

\subsection{Built-in support for ZeroChecks}
Many ZKP protocols use SumCheck to verify circuit correctness by checking that a polynomial $f(X)$ evaluates to zero across all inputs. However, simply verifying $\sum_X f(X) = 0$ is insufficient. Individual gates could evaluate to nonzero values (meaning they were incorrectly executed), but cancel each other out, so the total sum would still be zero. To prevent this, protocols like \cite{spartan, orion, strawman} use \textit{ZeroCheck}, which multiplies $f(X)$ with a randomized auxiliary polynomial $f_r(X)$
\footnote{In the ZKP literature \cite{spartan, hyperplonk, jolt}, $f_r(x)$ is written as $eq(x, r)$}
and checks that $\sum_X (f(X) \cdot f_r(X)) = 0$.
$f_r(X)$ is constructed on-the-fly from a vector of random challenges generated by a SHA module.
The construction of an $f_r(X)$ polynomial is referred to as the \textit{Build MLE} kernel in zkSpeed, and we integrate this auxiliary polynomial construction directly into the SumCheck datapath. During the first SumCheck round, one product lane is dedicated to computing $f_r(X)$. Its outputs are routed to a dedicated scratchpad bank and one EE, while the remaining EEs fetch MLE elements from the other scratchpad banks.
One less lane is available for product computations for the first SumCheck round, but this avoids separate $O(N)$ precomputation of $f_r$ and the memory overhead of writing and reading back values.
Subsequent SumCheck rounds treat $f_r$ as any other MLE fetched from and updated to off-chip.

\section{zkPHIRE Architecture}
We include the programmable SumCheck unit as part of zkPHIRE, a larger accelerator framework for the HyperPlonk protocol. \autoref{fig:overall_arch} shows the overall architecture.

\subsection{HyperPlonk Protocol Overview}
HyperPlonk has a very complex dataflow as outlined in~\cite{zkspeed2025}. Here, we describe the high-level protocol and which hardware modules each step requires. The 5 main steps are Witness Commitments, Gate Identity Check, Wire Identity Check, Batch Evaluations, and Polynomial Opening. 

In \textbf{Witness Commitments}, the prover commits to their private input values (witnesses) using multi-scalar multiplications (MSMs) so that they cannot claim other witness values later. We run this step with the MSM unit.
In \textbf{Gate Identity Checks}, the prover shows that all gates in the circuit were computed correctly via ZeroChecks. We run this step with our SumCheck unit and a Multifunction Forest unit.
In \textbf{Wire Identity Checks}, the prover proves that the wiring between gates is consistent with the circuit's intended connections
by enforcing a permutation argument over the wire values. This is done by constructing wiring permutation polynomials and forming a quotient and \textit{grand product} 
argument
that encodes the permutation constraint. These polynomials are committed to, and then correctness of the grand product is verified via a PermutationCheck (PermCheck).
We run this step with a Permutation Quotient Generator unit,
the SumCheck and Multifunction Forest units, and the MSM unit.

In \textbf{Batch Evaluations}, the prover evaluates all committed polynomials at various points provided by the verifier. We run this step with the Multifunction Forest unit.
In the \textbf{Polynomial Opening}, the prover reveals the values of multiple committed polynomials at specific points and proves that it is consistent with the original commitment. The prover then runs a SumCheck to combine these polynomial evaluation claims into one (we call this \textit{OpenCheck} as per zkSpeed).
The combined polynomial commitment is then opened using the MSM unit.
Overall this step uses an MLE Combine unit, the SumCheck and Multifunction Forest units, and the MSM unit.

\textbf{Masking ZeroCheck:} In zkSpeed and the HyperPlonk baseline, Gate Identity and Wire Identity steps run serially, despite verifying independent circuit properties. zkPHIRE includes support to overlap the Gate Identity’s ZeroCheck with the MSMs during Wiring Identity. MSMs dominate runtime and have low bandwidth pressure due to high data reuse~\cite{szkp, zkspeed2025}, so this scheduling effectively masks ZeroCheck latency.

\subsection{Accelerator Units}
\subsubsection{\textbf{SumCheck Unit with Sparsity Handling}} ZKP protocols often exhibit significant sparsity: enable MLEs ($q_i$) are mostly binary, while constant and witness MLEs ($q_c$, $w_i$) are typically 90\% sparse~\cite{szkp, pipezk, zkspeed2025}. zkPHIRE extends the SumCheck unit to exploit this sparsity for improved efficiency.

zkSpeed opts to use large on-chip scratchpads to store these MLEs, because they are reused throughout steps of the HyperPlonk protocol.
However, as the problem size scales, the area consumed by the scratchpad grows, with 10s of MBs dedicated to address translation buffers alone, and a total on-chip storage of nearly 300 MB.
Furthermore, the MLEs are only useful for round 1 of SumChecks. For subsequent rounds, the MLEs are updated and written to off-chip memory, so the global scratchpad is idle for roughly 50\% of the SumCheck runtime.
zkPHIRE opts to use smaller scratchpad buffers to store tiles of each MLE.
While this results in more access to off-chip memory and incurs more fill/drain latency penalties, it allows for more area to be dedicated to core computational structures. The improved speedups offset the latency hits incurred by using smaller scratchpads. 

zkSpeed's scratchpads compress MLEs but require address translation units to decode when MLE entries are sparse or dense. zkPHIRE eliminates the address translation overhead by fetching \textit{per-tile offset buffers} for witness MLEs.
For a tile of data fetched from off-chip, the SumCheck controller uses the offset buffer to determine where 255-bit elements begin in the bitstream; all other bits are interpreted as $0/1$ MLE entries and written into the corresponding addresses in the scratchpads.
The enable MLEs are stored as-is without using address translation. 
These optimizations incur minimal bandwidth costs while enabling more compact storage.

\subsubsection{\textbf{Multifunction Forest}}
In HyperPlonk, several sub-steps naturally map to binary-tree computations, such as MLE evaluation, computing the product MLE ($\pi$), and building MLEs~\cite{mo2025mtu}. 
We adopt the same base tree architecture as zkSpeed's MTU~\cite{mo2025mtu, zkspeed2025}, but differ in how it is integrated.
Whereas zkSpeed uses a single tree unit alongside a dedicated SumCheck module, zkPHIRE instantiates multiple tree units as a Multifunction Forest.
These units are reused to implement product-lane computations, removing the need for the extra modular multipliers in zkSpeed’s SumCheck module while keeping the pipeline fully utilized.
This shared design allows the same multipliers to serve both SumCheck products and other tree-based operations, improving utilization and reducing area.
As a result, zkPHIRE achieves the same latency as zkSpeed for the same workload with 15\% fewer multipliers.

\subsubsection{\textbf{MSM}}
We use the same MSM architecture as zkSpeed, since we target the same protocol, HyperPlonk. In HyperPlonk, the number of MSMs only depends on the number of witnesses. The underlying hardware is unchanged, we just compute more MSMs using it. Jellyfish gates have 5 witnesses, so we run 5 Sparse MSMs and 3 Dense MSMs.

\subsubsection{\textbf{MLE Combine}}
The MLE Combine module, shown in \autoref{fig:overall_arch}, is used for various element-wise operations and dot products before and after the OpenCheck in Polynomial Opening. For these steps, we fetch MLE tables from off-chip into up to 6 local SRAM buffers and perform the operations between MLE entries and/or challenges from prior protocol steps stored in registers. All operations are fully pipelined.

\subsubsection{\textbf{Permutation Quotient Generator}}

\begin{figure}[t!]
    \centering
    \includegraphics[width=\columnwidth]{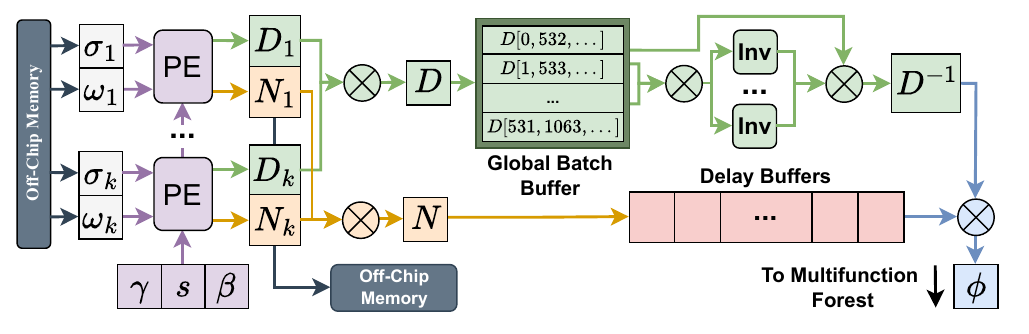}
    \caption{The Permutation Quotient Generator Module.}
    \label{fig:genND_fracMLE}
\end{figure}
PermCheck requires constructing Numerator ($N$), Denominator ($D$), Fraction ($\phi$), and Product ($\pi$) MLEs \cite{zkspeed2025}. Since $\phi$ is computed on-the-fly from N and D, we design a pipelined unit that generates all three simultaneously, producing one element per cycle after warmup (\autoref{fig:genND_fracMLE}). The Multifunction Forest is used to construct $\pi$ from $\phi$, as in zkSpeed. To accommodate both Vanilla and custom gates, we allocate 5 PEs (one per witness), with the ability to support more than 5 through overlapped scheduling and cyclic reuse of PEs. Intermediate $N_{1\dots k}$ and $D_{1\dots k}$ MLEs are written to HBM and combined with modular multiplications to produce the final $N$ and $D$ MLEs, followed by modular inversion of $D$.

We build on zkSpeed’s Montgomery batching approach~\cite{Montgomery_1987}, which uses a batch size of 64 and dedicated multipliers to mask inversion latency at high area cost. To improve efficiency, we reduce the batch size to 2 and eliminate per-inverse multipliers, instead using two shared multipliers, one for batching and one for output isolation. Since zkSpeed's modular multipliers are 17.7$\times$ larger than their inverse units (0.478 mm$^2$ vs. 0.027 mm$^2$ in TSMC 22nm), we increase the number of inverse units and schedule them in a round-robin fashion to initiate one inversion every two cycles without backpressure. We find that 266 inverse units is sufficient for this. A shared batch buffer stores inputs; once an inverse completes, it is multiplied by its paired batch element and the slot is reused. Our design achieves a 4.2$\times$ area reduction over zkSpeed without sacrificing throughput.

\subsubsection{\textbf{Interconnect and Memory}}
HyperPlonk’s stage-level data obliviousness enables static scheduling of computation and communication via a centralized controller. 
zkPHIRE comprises six major modules connected by a multi-channel shared bus, provisioned to meet peak data movement demands. 
During Wiring Identity, where bidirectional transfers occur between the 
SumCheck unit and Multifunction Forest, and one-way from Permutation Quotient Generator to MSM unit, up to three channels are needed to avoid stalls.
At our 294mm$^2$ design point, the peak bandwidth requirement reaches 19~TB/s, comparable to state-of-the-art accelerators ~\cite{samardzic2021_F1, kim2022_BTS}.

Each module includes local SRAM scratchpads to buffer intermediate data. Smaller buffers (6 MB) serve the Permutation Quotient Generator, MLE Combine, and the Multifunction Forest modules. Heftier modules—MSM (43~MB) and SumCheck (6~MB)—use highly banked SRAM to support multiple PEs and exploit data reuse. For instance, MSM stores elliptic curve points, while SumCheck stores frequently accessed MLE tiles. All modules connect to off-chip PHYs via the shared bus, coordinated to avoid contention and sustain peak throughput.

\section{Methodology}
\label{sec:Methodology}
We use Catapult HLS 2024 to generate the RTL for Montgomery multipliers with arbitrary primes (as in prior work~\cite{szkp, zkspeed2025}) and fixed primes~\cite{BLS12_381_paper, BLS12_381_web}, fully-pipelined PADD units, modular inverse units, and the SumCheck PEs.
We use the same elliptic curve as HyperPlonk (BLS12-381), where all MLE datatypes (e.g., in the SumCheck unit) are 255-bit, and all elliptic curve points (e.g., in the PADD) are 381-bit.
For logic synthesis, we use Design Compiler with TSMC 22nm. 
We explore the use of both arbitrary and fixed-prime modular multipliers (as done in \cite{nocap}); the latter saves us roughly 50\% on area and improves our computational density by roughly $2\times$.
Our 255b modular multipliers are 0.478/0.264 mm$^2$ (arbitrary/fixed); our 381b modular multipliers are 1.13/0.582 mm$^2$ (arbitrary/fixed).
We use Synopsys 22nm Memory Compiler for SRAM estimation.
For SHA3, we use an IP block from OpenCores \cite{opencores_sha3}.
To match prior work, we use scale factors of $3.6$ for area and $3.3$ for power to scale to 7nm ~\cite{szkp, zkspeed2025, 22nm_TSMC_web, 16nm_TSMC_paper, 7nm_TSMC_paper}.
We use a 1GHz clock for zkPHIRE.
For workload and sparsity, we use the same statistics from prior work~\cite{szkp, gzkp, pipezk, libsnark}.

The HyperPlonk CPU performance is measured on an AMD EPYC 7502 32-core processor running up to 3.35GHz, and the total die size is 296 mm$^2$~\cite{cpu_amd1, cpu_amd2, cpu_amd3}. 
We benchmark the GPU performance on an NVIDIA A100 GPU with 40GB memory and 1.6 TB/s bandwidth~\cite{NVIDIA-A100-40-GPU}.
We benchmark our SumCheck unit performance relative to prior work \cite{zkspeed2025} and a 4-threaded CPU implementation.
We then use workloads from prior work \cite{hyperplonk, libsnark} to model the performance of our full architecture vs. 32 CPU threads, up to problem sizes of $2^{30}$ nominal gates. We show these results in Section~\ref{sec:Evaluation}.

We use HLS tools to simulate core modules (modular multipliers, product lanes, extension engines, etc.) and use HLS-generated RTL to extract exact cycle counts, initiation intervals, and pipeline depths.
We use cycle-accurate simulators with analytical models of bandwidth constraints to estimate runtimes for MSMs and for the traversal-dependent behavior of the Multifunction Forest \cite{mo2025mtu}.
For SumCheck, we integrate HLS-generated cycle counts into a higher-level model (the automated scheduler in \autoref{fig:scheduler}) and then simulate runtimes analytically (accounting for off-chip bandwidth).
This approach is similar to prior work on accelerating cryptographic computing \cite{zkspeed2025, szkp, samardzic2021_F1, craterlake}.

\section{Evaluation}
\label{sec:Evaluation}
We present our evaluation in two parts. First, we evaluate our programmable SumCheck unit as a standalone unit in \autoref{sec:standalone_sumcheck}. We use polynomial gates seen in prior works \cite{verifiable_asics, spartan} as well as gates used for elliptic curve addition in the Halo2 \cite{halo2} library. We then evaluate it as part of a broader HyperPlonk accelerator, zkPHIRE, in \autoref{sec:zkphire_eval}.

\begin{table}[t]
    \caption{List of polynomial constraints.}
    \label{tab:polynomials}
    \centering
    \begin{tabular}{|c|c|c|}
        \hline
        \textbf{ID} & \textbf{Name} & \textbf{Polynomial} \\
        \hline
        0 & Verifiable ASICs \cite{verifiable_asics} & $q_{\text{add}}\cdot(a + b) + q_{\text{mul}}\cdot(a \cdot b)$ \\
        \hline
        1 & Spartan 1 \cite{spartan} & $(A \cdot B - C) \cdot f_\tau$ \\
        \hline
        2 & Spartan 2 \cite{spartan} & $(\text{Sum}_{ABC}) \cdot Z$ \\
        \hline
        \multicolumn{3}{|c|}{\textbf{Halo2 Constraints \cite{halo2}}} \\
        \hline
        3 & Nonzero Point Check & $q_\text{point}^\text{non-id} \cdot (y^2 - x^3 - 5)$ \\
        \hline
        4 & $x$-gated Curve Check & $(q_\text{point} \cdot x) \cdot (y^2 - x^3 - 5)$ \\
        \hline
        5 & $y$-gated Curve Check & $(q_\text{point} \cdot y) \cdot (y^2 - x^3 - 5)$ \\
        \hline
        6 & Incomplete Addition 1 & \makecell[c]{$q_\text{add-incomplete} \cdot ((x_r + x_q + x_p) $ \\  $\cdot(x_p - x_q)^2 - (y_p - y_q)^2)$} \\
        \hline
        7 & Incomplete Addition 2 & \makecell[c]{$q_\text{add-incomplete} \cdot  (y_r + y_q)(x_p - x_q)$ \\ $- (y_p - y_q)(x_q - x_r)$} \\
        \hline
        8 & Complete Addition 1 & \makecell[c]{$q_\text{add} \cdot (x_q - x_p)((x_q - x_p)\lambda$ \\ $- (y_q - y_p))$} \\
        \hline
        9 & Complete Addition 2 & \makecell[c]{$q_\text{add} \cdot (1 - (x_q - x_p)\alpha)$ \\ $(2y_p\lambda - 3x_p^2)$} \\
        \hline
        10 & Complete Addition 3 & \makecell[c]{$q_\text{add} \cdot x_p x_q (x_q - x_p)$ \\ $(\lambda^2 - x_p - x_q - x_r)$} \\
        \hline
        11 & Complete Addition 4 & \makecell[c]{$q_\text{add} \cdot x_p x_q (x_q - x_p)$ \\ $(\lambda(x_p - x_r) - y_p - y_r)$} \\
        \hline
        12 & Complete Addition 5 & \makecell[c]{$q_\text{add} \cdot x_p x_q (y_q + y_p)$ \\ $(\lambda^2 - x_p - x_q - x_r)$} \\
        \hline
        13 & Complete Addition 6 & \makecell[c]{$q_\text{add} \cdot x_p x_q (y_q + y_p)$ \\ $(\lambda(x_p - x_r) - y_p - y_r)$} \\
        \hline
        14 & Complete Addition 7 & $q_\text{add} \cdot (1 - x_p \cdot \beta)(x_r - x_q)$ \\
        \hline
        15 & Complete Addition 8 & $q_\text{add} \cdot (1 - x_p \cdot \beta)(y_r - y_q)$ \\
        \hline
        16 & Complete Addition 9 & $q_\text{add} \cdot (1 - x_q \cdot \gamma)(x_r - x_p)$ \\
        \hline
        17 & Complete Addition 10 & $q_\text{add} \cdot (1 - x_q \cdot \gamma)(y_r - y_p)$ \\
        \hline
        18 & Complete Addition 11 & \makecell[c]{$q_\text{add} \cdot (1 - (x_q - x_p)\alpha$ \\ $- (y_q + y_p)\delta)x_r$} \\
        \hline
        19 & Complete Addition 12 & \makecell[c]{$q_\text{add} \cdot (1 - (x_q - x_p)\alpha$ \\ $- (y_q + y_p)\delta)y_r$} \\
        \hline
        \multicolumn{3}{|c|}{\textbf{HyperPlonk (HP) Polynomials \cite{hyperplonk} }} \\
        \hline
        20 & Vanilla ZeroCheck & \makecell[c]{$(q_Lw_1 + q_Rw_2 - q_Ow_3$ \\$+ q_Mw_1w_2 +q_C)f_r$} \\
        \hline
        21 & \makecell[c]{Vanilla PermCheck \\ ($\alpha$ is a scalar)} & \makecell[c]{
        $(\pi - p_1p_2 + \alpha(\phi D_1D_2D_3$ \\ $ - N_1N_2N_3)) f_r$} \\
        \hline
        22 & Jellyfish ZeroCheck & \makecell[c]{$(q_1 w_1 + q_2 w_2 + q_3 w_3 + q_4 w_4 $ \\ $ 
      + q_{M_1} w_1 w_2 + q_{M_2} w_3 w_4 $ \\ $+ q_{H_1} w_1^5 + q_{H_2} w_2^5 + q_{H_3} w_3^5$ \\  $+ q_{H_4} w_4^5 - q_O w_5 $\\$+ q_{ecc} w_1 w_2 w_3 w_4 + q_C) f_r$} \\
        \hline
        23 & \makecell[c]{Jellyfish PermCheck \\ ($\alpha$ is a scalar)}  & \makecell[c]{
        $(\pi - p_1p_2 + \alpha(\phi D_1D_2D_3 \cdot $ \\ $D_4D_5 - N_1N_2N_3N_4N_5)) f_r$} \\
        \hline
        24 & OpenCheck & \makecell[c]{$
        y_1f_{r_1} + y_2f_{r_2} +y_3f_{r_3} $ \\ $y_4f_{r_4} + y_5f_{r_5} +y_6f_{r_6} $} \\
        \hline
    \end{tabular}
\end{table}

\begin{figure}[t!]
    \centering
    \hspace*{-5mm} % shift left
    \includegraphics[width=1.08\columnwidth]{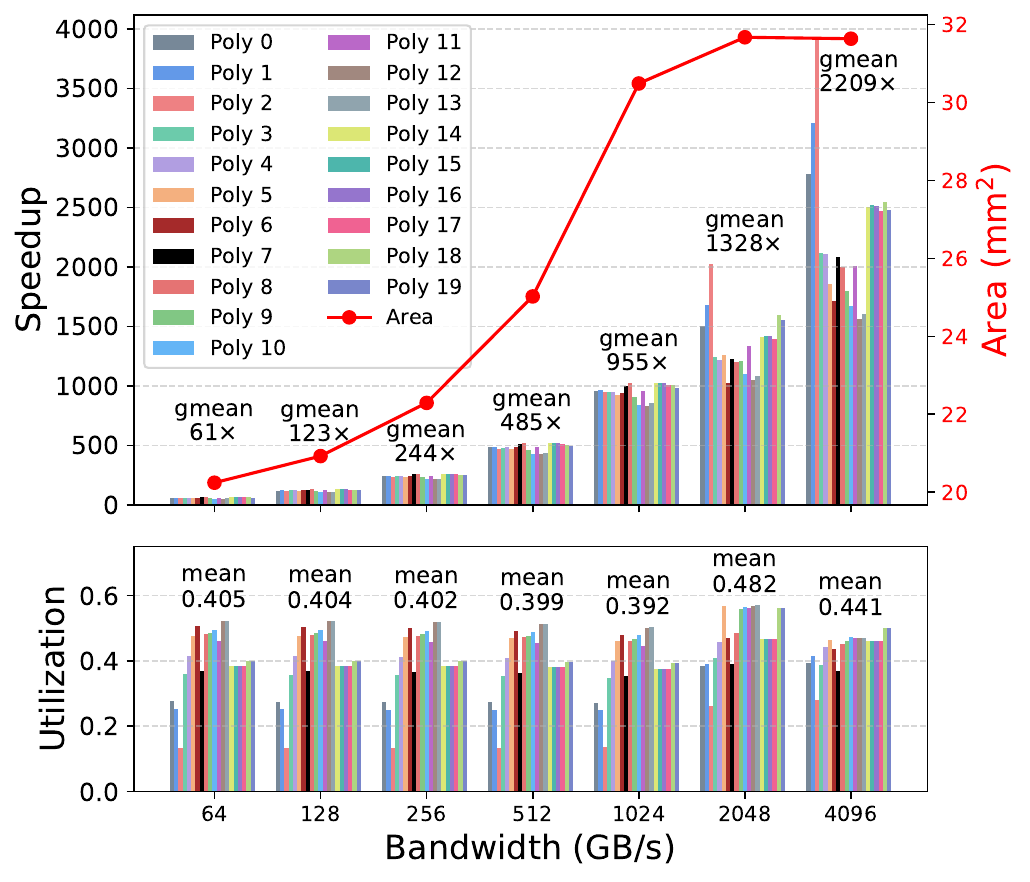}
    \caption{Speedups over 4-threaded CPU implementations of SumChecks on the polynomials in \autoref{tab:polynomials}. We pick a different optimal design across each bandwidth technology. Utilization across the design points is also shown.}
    \label{fig:cpu_speedup_and_util}
\end{figure}

\subsection{Programmable SumCheck}
\label{sec:standalone_sumcheck}

\subsubsection{Speedup over various gate types}
\label{sec:speedup_and_util}
\autoref{tab:polynomials} enumerates the polynomials we use in our evaluation. We first benchmark the performance of SumChecks on each polynomial by extending HyperPlonk's existing code base \cite{hyperplonk}. We then run each polynomial on our programmable SumCheck unit. In this evaluation, we seek to optimize both performance and utilization across all polynomials. We can always allocate EEs and PLs to handle the worst-case polynomial degree, however, this would lead to underutilization on low-degree polynomials. In our design space search, we specify an area and bandwidth constraint and then use the following objective function:
\begin{equation*}
    \min_{\text{design}} \; (1 - \lambda) \cdot f_{\text{slowdown}}(s_{d,i}) + \lambda \cdot (1 - f_{\text{util}}(u_{d,i}))
\end{equation*}
Here, $s_{d,i}$ refers to the slowdown of the $i^{th}$ polynomial for a given design $d$, relative to the fastest runtime for that polynomial in the area-constrained design space. Similarly, $u_{d,i}$ reflects the utilization for design $d$ on polynomial $i$. $f_\text{slowdown}$ and $f_\text{util}$ are aggregation functions; we employ geomean for slowdown and arithmetic mean for utilization. $\lambda$ controls the trade-off between the minimizing slowdown and maximizing utilization. For this analysis, we focus on utilization, using $\lambda = 0.8$. We compare the runtime against the CPU running each polynomial's SumCheck on 4 threads, for which we estimate the total core area to be 37 mm$^2$ in 7nm \cite{cpu_amd1, cpu_amd2, cpu_amd3} and use that as our area constraint (after scaling from 22nm). Across various bandwidths, we find the optimal design point and report speedups and utilizations in Figure \ref{fig:cpu_speedup_and_util}. For most designs, 2 EEs and 5 PLs is optimal.

We observe $\sim$ 50\% utilization across benchmarks at each bandwidth, due to three main factors: (1) MLE Update units are inactive during the first round of SumCheck, which accounts for about half of total runtime; (2) lower-degree polynomials require fewer product lanes; and (3) repeating MLEs (e.g., $q_{add}$) reduce demand on the update modular multipliers since they are reused across terms.
Consequently, when optimizing for utilization, we observe a tendency to limit the number of EEs. Further parallelism and potentially higher utilization can be achieved by mapping multiple terms to EEs when possible, but it would introduce additional complexity in interconnect and increase bandwidth demand. These trade-offs can be explored further in future work.
In contrast, product lane modular multipliers are usually fully utilized, provided the polynomial degree and memory bandwidth are high enough.
Despite moderate utilization, we get significant speedups over CPU baselines, approaching $1000\times$ at 1 TB/s of bandwidth. 

Our SumCheck unit was benchmarked on a ``training set" of mostly lower-degree polynomials, which comprise most ZKP protocols today. However, we found that higher-degree (e.g., Jellyfish polynomials) achieved comparable or higher utilization ($\approx 47\%$) than many lower-degree polynomials. This is because additional constituent polynomials and extension products place greater concurrent demand on EEs and product lanes, indicating that the utilization improves with greater polynomial complexity for a given hardware configuration.

\subsubsection{High-degree sweep}

\begin{figure}[t!]
    \centering
    \includegraphics[width=\columnwidth]{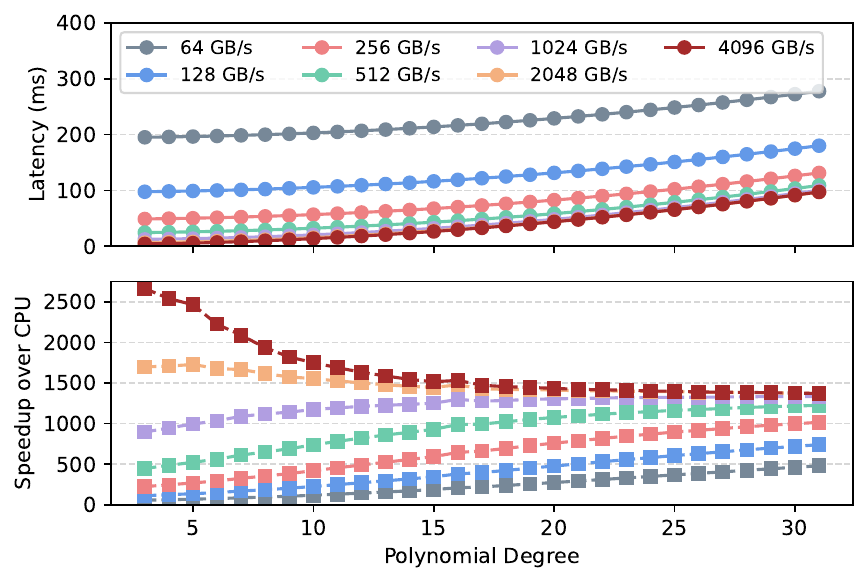}
    \caption{Performance of a fixed SumCheck configuration on high-degree polynomials at different bandwidths. Speedup relative to CPU is also plotted.}
    \label{fig:high_degree_sweep_bw_plot}
\end{figure}

We analyze the performance scaling of SumChecks as we increase polynomial degree. We sweep polynomial degrees ($d = 2\ldots30$) for $f = q_1w_1 + q_2w_2 + q_3w_1^{d-1}w_2 + q_c$, as in HyperPlonk, and measure speedup over CPU. We use the same objective as before and pick a high-performance design under the same area constraints (\autoref{fig:high_degree_sweep_bw_plot}). Our results show that low-degree polynomials require HBM-scale bandwidths to achieve $1000\times$ speedups, while higher degree polynomials can reach similar speedups with DDR5-level bandwidths (e.g., 256 GB/s). Higher degree polynomials in this scenario exhibit more computations on the same amount of data, which is why the disparity in speedups across bandwidths is less compared to low-degree polynomials.

\begin{figure}[t!]
    \centering
    \includegraphics[width=\columnwidth]{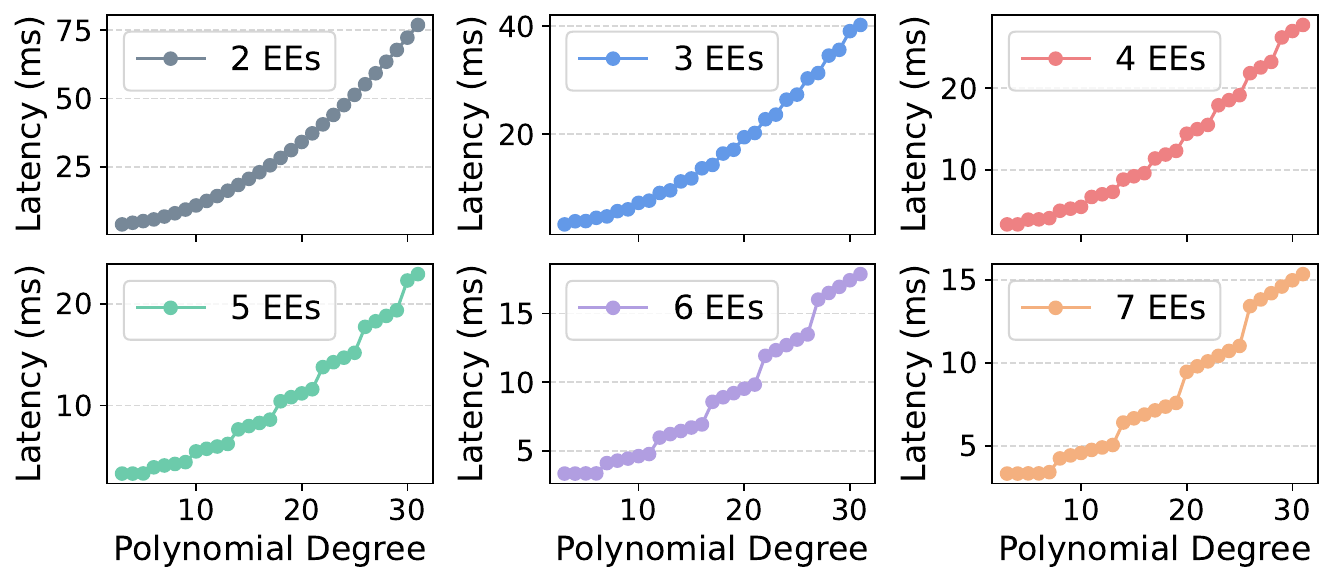}
    \caption{Scheduler-induced SumCheck behavior at fixed BW and PL settings.}
    \label{fig:ee_sweep}
\end{figure}

At fixed bandwidth, PE count, and EE-per-PE settings, increasing the number of PLs per PE improves performance roughly proportionally. Similar gains are observed when increasing EE count at fixed PLs. Interestingly, as shown in \autoref{fig:ee_sweep}, runtime increases in discrete jumps due to the graph decomposition the scheduler uses to map polynomials to EEs (see \autoref{fig:scheduler}). For example, under 6 EEs, degree-$1$–$6$ polynomials have 1 node in their schedule graph, while degree-$7$–$11$ polynomials require 2. Each added node incurs a sharp runtime jump, as it introduces additional products across all MLEs and terms. Within each node cluster, runtime grows more gradually with degree due to early exit optimizations. These trends may help optimize the SumCheck unit when the degree distribution of polynomials is known, by minimizing high-degree outliers that cause runtime spikes.

\subsubsection{Comparison with Prior ASICs}
\label{sec:asic_comp_sumcheck}
\begin{figure}[t!]
    \centering
    \includegraphics[width=\columnwidth]{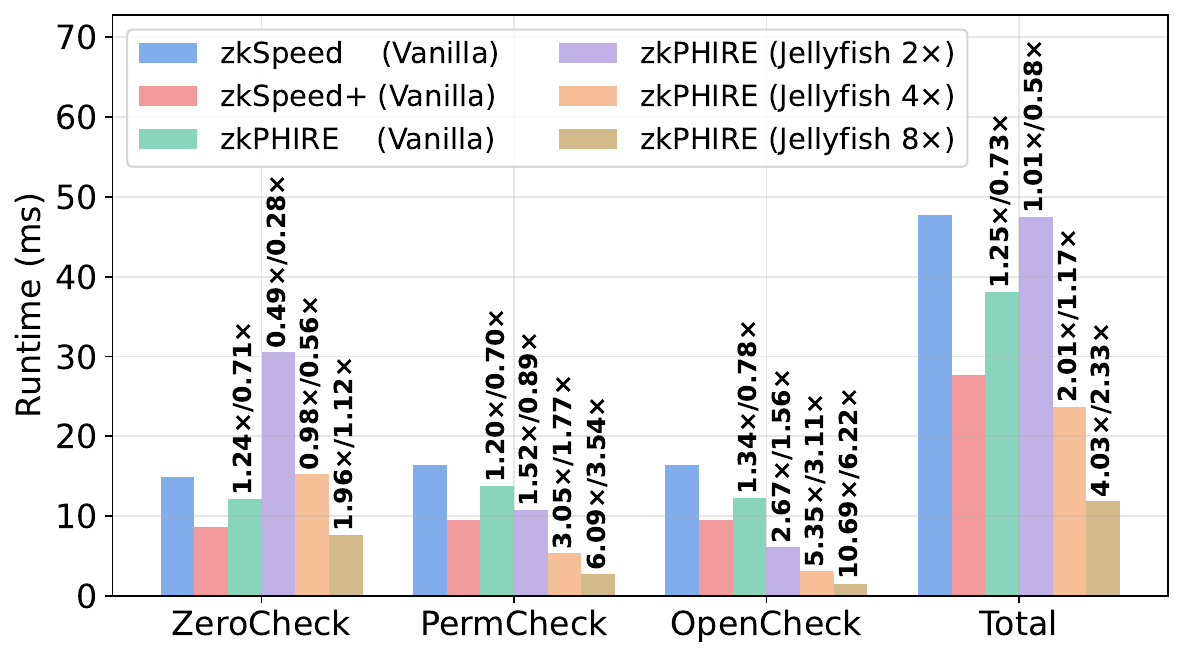}
    \caption{Comparison with prior ASICs. Bars are annotated with speedup numbers over baseline zkSpeed / zkSpeed+. }
    \label{fig:asic_speedups}
\end{figure}

\autoref{fig:asic_speedups} details our speedups relative to a prior ASIC, zkSpeed. zkSpeed implements the SumChecks by using a custom unified core that maximizes immediate reuse. We also compare with zkSpeed+, which gives zkSpeed the benefits of pipelining MLE updates directly into the extensions and product computations (as we do in zkPHIRE). We assume the same modular multiplier area and 2 TB/s as zkSpeed. We use the same objective function from \autoref{sec:speedup_and_util} on the same training set of polynomials in \autoref{tab:polynomials} with a roughly iso-zkSpeed area constraint. We pick a 35.24 mm$^2$ design compared to zkSpeed's 30.8 mm$^2$ SumCheck and MLE Update area. Our design has 4.9 mm$^2$ of SRAM, while zkSpeed used a much larger global SRAM for storing MLEs, hence we believe this comparison is fair.

We run the same polynomials from zkSpeed's evaluation on our SumCheck unit, and find that compared to the zkSpeed+, we are only 30\% slower at iso-area and iso-bandwidth, while offering programmability to support other types of polynomials beyond those in HyperPlonk. We then use the same hardware design point for evaluation on Jellyfish polynomials. The advantage of using Jellyfish gates is that for a given application, while the polynomial complexity grows, the total gate count (i.e., workload length) decreases. We simulate three degrees of gate count reduction, $2\times, 4\times, \text{and } 8\times$.

We find that, for ZeroCheck, because the Jellyfish polynomial is more complex, the $2\times$ is not sufficient to bring speedups over the Vanilla counterpart. However, PermCheck's increase in complexity \textit{is} outweighed by a $2\times$ workload reduction. OpenCheck, which is the same between Jellyfish and Vanilla, exhibits $1.12\times, 3.54\times$ and $6.22\times$ speedup for each gate reduction over zkSpeed+. 
Combining all three SumChecks, a Jellyfish reduction by $4\times$ is sufficient to outperform Vanilla polynomials on both zkSpeed+ and zkPHIRE. These gains are observed in spite of the inherent overheads of programmability in our accelerator.

\subsubsection{Comparison with GPU}
\label{sec:gpu_comparison}
We evaluate zkPHIRE against the GPU library ICICLE~\cite{icicle} by running SumChecks from the Spartan and HyperPlonk protocols on an NVIDIA A100 GPU. 
We use the same SumCheck design from \autoref{sec:asic_comp_sumcheck}, assuming 1 TB/s of bandwidth to roughly match the A100.
Spartan is chosen due to its adoption in several recent protocols~\cite{jolt, kwan2024verifying, nova}, while HyperPlonk evaluates performance on high-degree gates.
For context, we also report CPU runtimes for these SumChecks in \autoref{tab:gpu_cpu_zkphire}.
Notably, ICICLE currently supports composite polynomials with at most eight unique constituent polynomials, preventing evaluation of polynomials 21-24 in \autoref{tab:polynomials}. We choose to evaluate the Vanilla Gate portion of polynomial 20 (excluding $f_r$).
We find that zkPHIRE accelerates these SumChecks by roughly $70\times$ over GPU, and by $600-800\times$ over CPU.

\begin{table}[t!]
\centering
\caption{SumCheck Runtimes on CPU, GPU, and zkPHIRE for $N=24$. HyperPlonk polynomials are from \autoref{tab:polynomials}. zkPHIRE runtimes are accompanied with speedups over CPU/GPU}
\label{tab:gpu_cpu_zkphire}
\resizebox{\columnwidth}{!}{
\begin{tabular}{|c|c|c|c|c|c|}
\hline
\centering\textbf{Polynomial} & 
\centering\textbf{Number of } & 
\centering\textbf{Problem} & 
\multicolumn{3}{c|}{\textbf{Runtime (ms)}} \\ \cline{4-6}
\textbf{Type} & \textbf{SumChecks} & \textbf{Size} & \textbf{CPU} & \textbf{GPU} & \textbf{zkPHIRE} \\ \hline \hline\textbf{$(A \cdot B - C) \cdot f_\tau$} & 1  & $2^N$     & 6770  & 571  & 7.6 ($891\times$/$75\times$) \\ \hline
$(\text{Sum}_{ABC}) \cdot Z$   & 1  & $2^{N+1}$ & 5237  & 586  & 8.4  ($623\times$/$70\times$) \\ \hline
$A\cdot B \cdot C$ & 12 & $2^N$     & 60993 & 5376 &             78.9 ($773\times$/$68\times$) \\ \hline
$A\cdot B \cdot C$ & 6  & $2^{N-1}$ & 15248 & 1440 &             19.7  ($774\times$/$73\times$) \\ \hline
$A\cdot B \cdot C$ & 4  & $2^{N+1}$ & 40662 & 3460 &             52.6 ($773\times$/$66\times$) \\ \hline \hline

HP Poly. 20 \tablefootnote{$f_r$ polynomial excluded} & 1  & $2^N$ & 13354 & 1089 &  15.8 ($845\times$/$69\times$) \\ \hline
HP Poly. 21 & 1  & $2^N$ & 21625 & --  &  22.7 $(953\times)$ \\ \hline
HP Poly. 22 & 1  & $2^N$ & 74226 & --  &  69.5 $(1068\times)$ \\ \hline
HP Poly. 23 & 1  & $2^N$ & 32774 & --  &  32.2 $(1018\times)$ \\ \hline
HP Poly. 24 & 1  & $2^N$ & 17591 & --  &  21.3 $(826\times)$ \\ \hline
\end{tabular}
 }
\end{table}

\subsection{zkPHIRE}
\label{sec:zkphire_eval}

We now incorporate the programmable SumCheck unit into a larger accelerator that accelerates the HyperPlonk protocol, zkPHIRE. Here, our analysis focuses on yielding high-performance designs. In these comparisons, we assume the CPU runs with 32 threads, roughly 296 mm$^2$ in 7nm. These experiments assume fixed-prime modular multipliers, which is a common technique seen in prior work \cite{nocap, hyperplonk, libsnark}.
\begin{table}[t!]
\centering
\caption{Design Space Exploration Parameters for zkPHIRE.}
\label{tab:design_space}
\resizebox{\columnwidth}{!}{
\setlength{\tabcolsep}{2mm}{
\begin{tabular}{|c|c|c|}
\hline
\textbf{Module} & \textbf{Design Knob} & \textbf{Values}   \\ \hline \hline
\textbf{SumCheck} & PEs  & 1, 2, 4, 8, 16, 32\\ \hline
\textbf{SumCheck} & Extension Engines  & 2, 3, 4, 5, 6, 7 \\ \hline
\textbf{SumCheck} & Product Lanes  & 3, 4, 5, 6, 7, 8 \\ \hline
\textbf{SumCheck} & SRAM Bank Size  & $2^{10}-2^{15}$ \\ \hline

\textbf{MSM} & PEs                & 1, 2, 4, 8, 16, 32    \\ \hline
\textbf{MSM} & Window Size       & 7, 8, 9, 10    \\ \hline
\textbf{MSM} & Points/PE & 1K, 2K, 4K, 8K, 16K \\ \hline
\textbf{FracMLE} & PEs & 1, 2, 3, 4 \\ \hline
\textbf{---} & Bandwidth (GB/s) & 64, 128, 256, 512, 1T, 2T, 4T \\ \hline
\end{tabular}
}}
\end{table}

\begin{figure}[t!]
    \centering
    \includegraphics[width=\columnwidth]{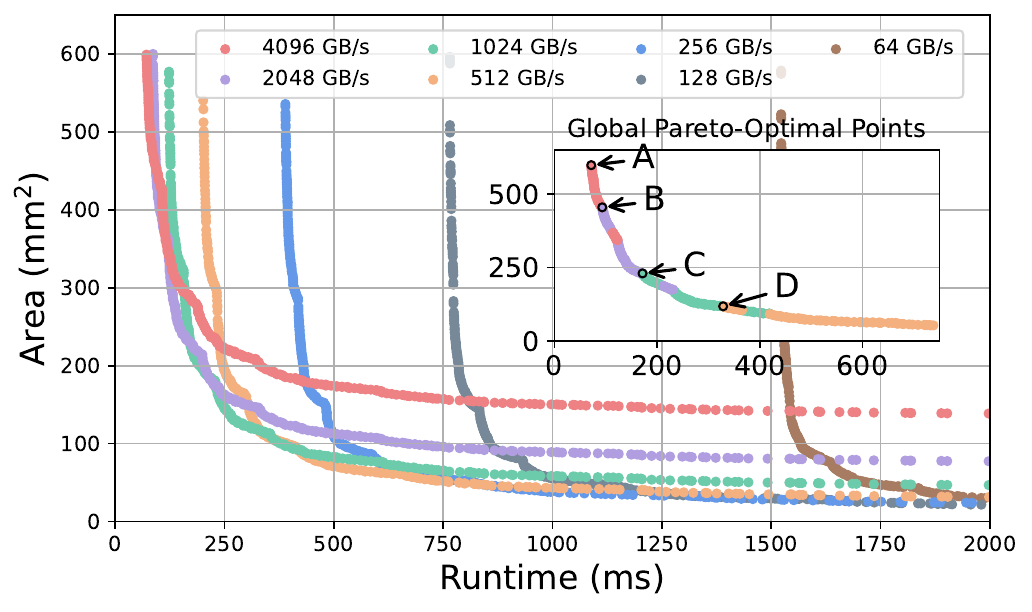}
    \caption{Pareto Frontiers for $2^{24}$ gates. We plot the individual Pareto curves for each bandwidth tier and the global Pareto frontier in the inset. We label the highest performing design for each bandwidth tier (A - D) in the inset.}
    
    \label{fig:pareto_plot}
\end{figure}

\begin{table}[t!]
\centering
\caption{Runtime, Area, and Bandwidth for Globally Pareto-optimal zkPHIRE Designs in Figure \ref{fig:pareto_plot}.}
\label{tab:pareto_table}
\resizebox{\columnwidth}{!}{
\setlength{\tabcolsep}{2mm}{
\begin{tabular}{|c|c|c|c|c|}
\hline
\textbf{Design} & \textbf{Runtime (ms)} & \textbf{Area (mm$^2$)} & \textbf{BW (GB/s)} & \textbf{CPU Speedup} \\ \hline \hline
\textbf{A}   & 71.436   & 599.08 & 4096 & 2560$\times$ \\ \hline
\textbf{B}   & 92.887   & 455.23 & 2048 & 1969$\times$ \\ \hline
\textbf{C}   & 171.332  & 229.72 & 1024 & 1067$\times$ \\ \hline
\textbf{D}   & 328.463  & 117.56 & 512  & 557$\times$ \\ \hline
\textbf{E}  & 477.377  & 75.14  & 512  & 383$\times$ \\ \hline
\textbf{F}  & 786.298  & 49.99  & 512  & 233$\times$ \\ \hline
\textbf{G} & 1716.765 & 25.03  & 128  & 107$\times$ \\ \hline
\end{tabular}
}}
\end{table}

\subsubsection{Pareto Frontier Analysis}

We sweep the parameters in \autoref{tab:design_space} to obtain Pareto-optimal designs for each bandwidth, then construct a global Pareto frontier across all bandwidths.
\autoref{fig:pareto_plot} illustrates the design space for a $2^{24}$ Jellyfish gate workload under seven bandwidth scenarios, ranging from DDR-class to HBM-scale bandwidths.
We account for memory PHY overheads~\cite{nocap, osiris, haac, hbm2, samardzic2021_F1, craterlake, ark, sharp}, assuming 14.9~mm\textsuperscript{2} per HBM2 PHY and 29.6~mm\textsuperscript{2} per HBM3 PHY~\cite{hbm3}.
For $2^{24}$ Jellyfish gates, the CPU runtime is roughly 182.896 seconds. We can hit roughly $1000\times$ speedup with a 207 mm$^2$ design and 1 TB/s off-chip bandwidth, and with $294$ mm$^2$ and 2 TB/s, around $1400\times$ at iso-CPU area.
For this high-performance regime, a Pareto design around 250~mm\textsuperscript{2} with 1 TB/s (design B) achieves nearly $2\times$ and $3\times$ speedup over 512~GB/s and 256~GB/s configurations. These results highlight a key trend in recent cryptographic accelerators where high performance demands investments in high-bandwidth memory technologies (even for small accelerators like NoCap, where the PHY cost alone is $3\times$ the compute area). That being said, with our accelerator framework, for applications with looser prover constraints, DDR-scale or lower bandwidths can still yield $100-200\times$ speedups over multi-threaded CPU baselines. In \autoref{tab:pareto_table}, we enumerate three additional design points; we can see that a two-order-of-magnitude speedup is achievable in under 100 mm$^2$, relatively small by crypto-ASIC standards \cite{kim2022_BTS, craterlake, ark, reagen2021cheetah}.

\begin{figure}[t!]
    \centering
    \includegraphics[width=\columnwidth]{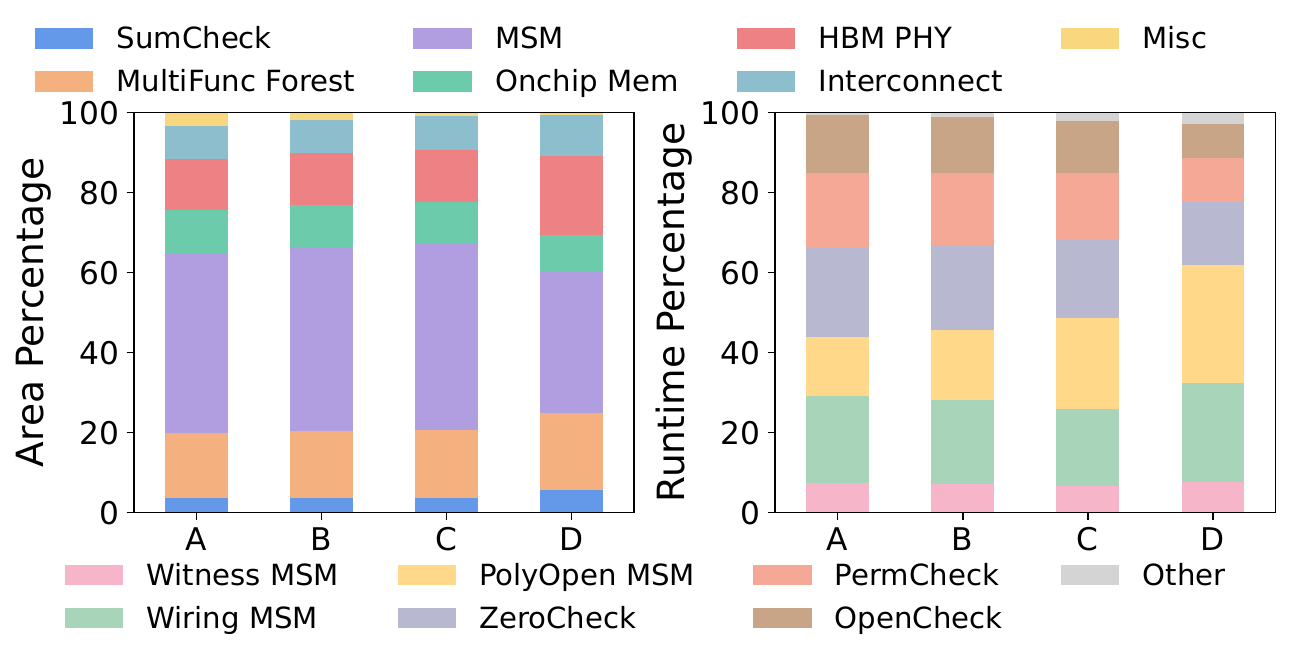}
    \caption{Area (left, top legend) and runtime (right, bottom legend) breakdowns for the selected Pareto points in \autoref{fig:pareto_plot}.}
    \label{fig:area_breakdown}
\end{figure}

\begin{figure}[t!]
    \centering
    \includegraphics[width=\linewidth]{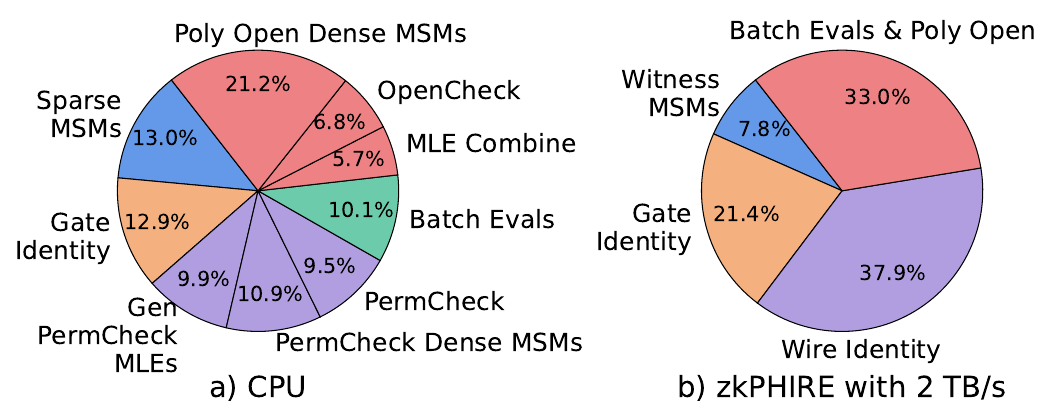}
    \caption{Runtime breakdown for CPU and zkPHIRE for $2^{24}$ Jellyfish gates. CPU's sequential execution enables finer breakdowns compared to zkPHIRE.}
    \label{fig:runtime_breakdown}
\end{figure}

\subsubsection{Area, Runtime, and Power}

In \autoref{fig:area_breakdown}, we analyze selected Pareto points from \autoref{fig:pareto_plot}. Across all points, MSM dominates the area as expected.
As the bandwidth increases, both MultiFunction Forest and MSM portions increase, but from C to D, the absolute MSM area remains unchanged while the SumCheck and Forest area increase.
This is also evident in the runtime breakdown: from C to D, the SumCheck portions (ZeroCheck, PermCheck, and OpenCheck) become smaller. Because SumCheck is memory-bound, higher bandwidth incentivizes more resource allocation towards SumCheck-related compute. For low-performance designs, less bandwidth is required, so the MSM is allocated a higher area portion.

\begin{table}
\caption{Area and power of zkPHIRE. Other includes the Permutation Quotient Generator, MLE Combine and SHA3 units.
}
\label{tab:all_area_power}
\footnotesize
\resizebox{0.98\columnwidth}{!}{
\setlength{\tabcolsep}{1mm}{
\begin{tabular}{lrr}
\hline
& \multicolumn{1}{l}{\textbf{Area (mm²)}} & \multicolumn{1}{l}{\textbf{Average Power (W)}} \\ \hline
MSM (32 PEs)                 & 105.69                                  & 58.99                                          \\
Multifunc Forest ({80} Trees)  & 48.18                                   & {40.69}                                          \\
SumCheck (16 PEs)            & 16.65                                   & 14.43                                          \\
Other                        & {10.64}                                    & {6.17}                                           \\ \hline
\textbf{Total Compute}       & \textbf{181.15}                         & {\textbf{120.29}}                                \\ \hline
SRAM                         & {27.55} & {3.56}                                \\
Interconnect                 & 26.42                                   & 14.83                                          \\
HBM3 (2 PHYs)                & 59.20                                   & 63.60                                          \\
\textbf{Total Memory System} & \textbf{{113.18}}                         & \textbf{{81.99}}                                 \\ \hline
\textbf{Total}               & \textbf{294.32}                         & \textbf{{202.28}}                                \\ \hline
\end{tabular}
}
}
\end{table}

Table~\ref{tab:all_area_power} shows an exemplar zkPHIRE design point we choose for the following analysis. 
In this 294 mm$^2$ design, there are 32 MSM PEs, with {8} modular multipliers per Multifunction tree PE, and 7 EEs and 5 PLs per SumCheck PE. 
The design runs on 2 TB/s HBM bandwidth. 
We use two 32$\times$32 bit-sliced crossbars~\cite{samardzic2021_F1, kim2022_BTS, passas2012crossbar} in the MSM and SumCheck unit, with a shared-bus to form the on-chip bandwidth (up to 19 TB/s) between modules. 
Figure~\ref{fig:runtime_breakdown} further breaks down the runtime to compare with HyperPlonk on CPU.
The colors indicate the same HyperPlonk step between CPU and zkPHIRE running at the configuration of \autoref{tab:all_area_power}. 
We show the latency breakdown for Gate and Wire Identities before performing ZeroCheck masking to show their initial runtime proportions. MSMs dominate runtime before and after  acceleration, though compared to zkSpeed's CPU baseline~\cite{zkspeed2025}, SumChecks take more runtime because the polynomials are complex, exerting more memory pressure on the CPU.

\subsubsection{The On-Chip Memory Trade-off} Since our design space exploration (DSE) sweeps per-MLE capacity (1K - 16K words), we can assess the performance impacts of data reuse via larger scratchpads. 
Our DSE found that performance–area Pareto-optimal designs consistently select SumCheck units with less SRAM in favor of higher compute parallelism (e.g., more PEs/EEs/PLs).
For example, from Figure 10, the Pareto front contains a 100 mm$^2$ design (376 ms runtime) with 8 MSM PEs, 4 SumCheck PEs (4 EEs, 7 PLs), and 4K sized SRAMs. 
Increasing PLs from 7 to 8, we get a 104 mm$^2$ design and 366 ms, which is still Pareto optimal.
If we instead increase the SRAM size to 16K, we get a 108 mm$^2$ design at 368 ms. 
While this improves performance, it is not Pareto optimal. 
Therefore, increasing scratchpad size does improve performance but may not offer the best performance per area.

\begin{figure}[t!]
    \centering
    
    \hspace*{-.7cm}
    \includegraphics[width=\columnwidth]{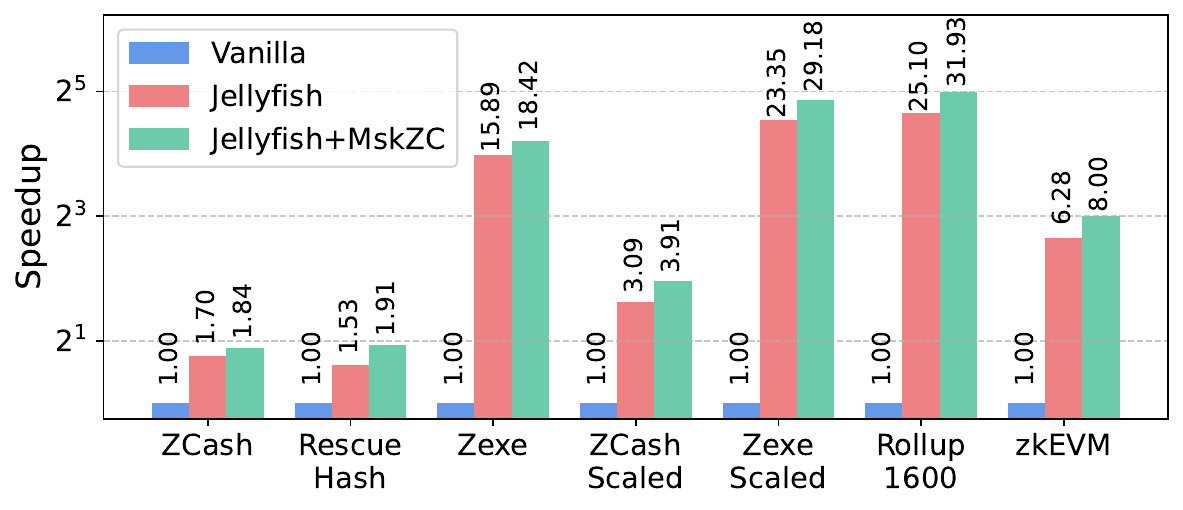}
    \caption{zkPHIRE speedups across workloads relative to Vanilla Gates. MskZC = Masked ZeroCheck optimization.}
    \label{fig:optimization_tests}
\end{figure}

\begin{figure}[t!]
    \centering    
    \hspace*{-.75cm}
    \includegraphics[width=\columnwidth]{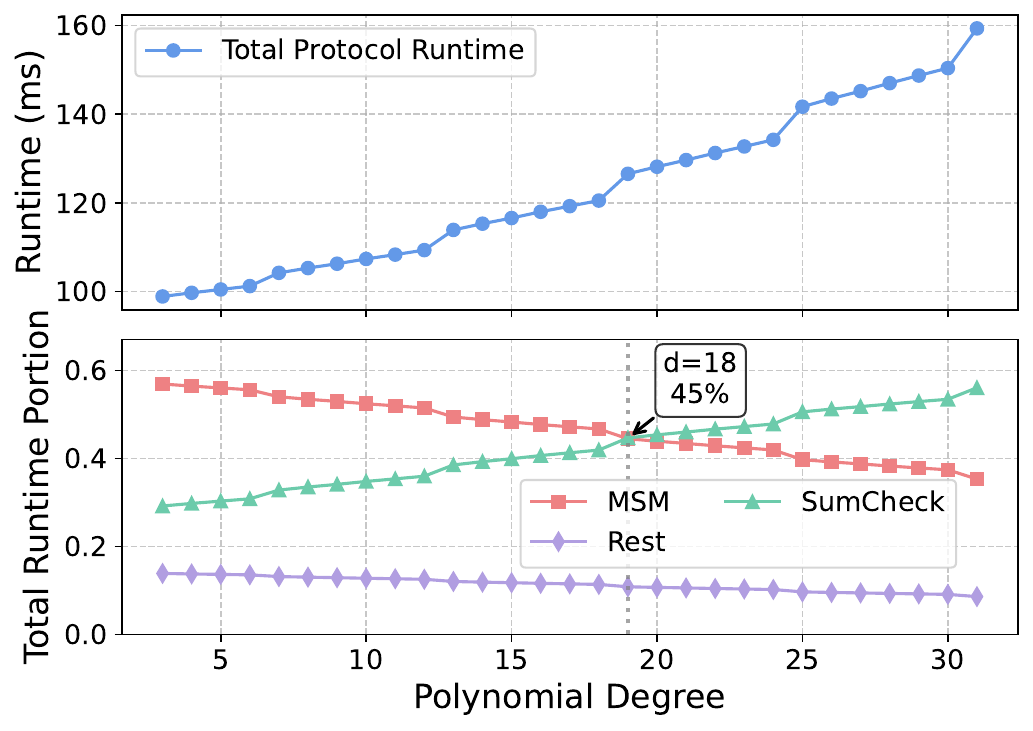}
    \caption{High-degree sweep on overall HyperPlonk Protocol. Crossover point is noted. $f = q_1w_1 + q_2w_2 + q_3w_1^{d-1}w_2 + qc$.}
    \label{fig:crossover}
\end{figure}

\subsubsection{Performance from Protocol-level Optimizations}
\autoref{fig:optimization_tests} quantifies the benefits of using Jellyfish gates and masking ZeroCheck latency across several workloads, since different workloads may respond differently to Jellyfish-style mappings. For large workloads, where the Jellyfish gate count is $> 2^{20}$, the speedups from Jellyfish alone reasonably reach what we expect from table size reduction alone. For smaller workloads, we see lower speedups from Jellyfish alone due to serialization overheads of MSMs and other off-chip memory overheads (e.g. fill/drain latencies) that are dwarfed for large workloads. To highlight this, we test scaled up versions of Zcash and Zexe to $2^{24}$ and $2^{25}$ (as done previously \cite{nocap}), and observe $3.1\times$ and $23.3\times$ speedups from Jellyfish alone. For zkEVM, no estimation of Vanilla gates is available \cite{hyperplonk}; we estimate an $8\times$ reduction to highlight our speedup trends for a hypothetical case. ZeroCheck masking provides additional gains of roughly 25-27\% for most workloads, which is what we expect from Amdahl's law by inspecting \autoref{fig:runtime_breakdown}. 

\subsubsection{High-Degree Sweep}
In \autoref{fig:crossover}, we plot the runtime of the highlighted design point over custom gates represented by $f = q_1w_1 + q_2w_2 + q_3w_1^{d-1}w_2 + qc$. For this family of gates, the number of witnesses is fixed, so the total MSM runtime is constant across all values of $d$. Consequently, there is a crossover point after which SumChecks dominate the total protocol runtime. This highlights a key design trade-off: while higher-degree gates reduce overall gate count, they can potentially shift the performance bottleneck from MSMs to SumChecks. Protocol designers must therefore balance gate expressivity with the rising cost of polynomial verification.

\begin{table}[]
\caption{Runtime Comparison with zkSpeed+ and CPU for Vanilla gates (speedups are relative to CPU).}
\label{tab:iso_zkspeed_comp}
\resizebox{\columnwidth}{!}{%
\begin{tabular}{|c|c|ccc|}
\hline
\multirow{2}{*}{Workload} & \multirow{2}{*}{Gates} & \multicolumn{3}{c|}{Runtime (ms)} \\ \cline{3-5} 
 &  & \multicolumn{1}{c|}{CPU} & \multicolumn{1}{c|}{zkSpeed+} & zkPHIRE \\ \hline
ZCash & 2\textsuperscript{17} & \multicolumn{1}{c|}{1429} & \multicolumn{1}{c|}{1.825 (783$\times$)} & 2.012 (710$\times$) \\ \hline
Auction & 2\textsuperscript{20} & \multicolumn{1}{c|}{8619} & \multicolumn{1}{c|}{10.171 (847$\times$)} & 10.88 (792$\times$) \\ \hline
2\textsuperscript{12} Rescue Hashes & 2\textsuperscript{21} & \multicolumn{1}{c|}{18637} & \multicolumn{1}{c|}{19.631 (888$\times$)} & 20.977 (1822$\times$) \\ \hline
Zexe's Recursive Ckt & 2\textsuperscript{22} & \multicolumn{1}{c|}{37469} & \multicolumn{1}{c|}{38.535 (972$\times$)} & 41.117 (911$\times$) \\ \hline
Rollup of 10 Pvt Tx & 2\textsuperscript{23} & \multicolumn{1}{c|}{74052} & \multicolumn{1}{c|}{76.356 (969$\times$)} & 81.362 (910$\times$) \\ \hline
Rollup of 25 Pvt Tx & 2\textsuperscript{24} & \multicolumn{1}{c|}{145500} & \multicolumn{1}{c|}{151.973 (957$\times$)} & 161.876 (898$\times$) \\ \hline
Rollup of 50 Pvt Tx & 2\textsuperscript{25} & \multicolumn{1}{c|}{325048} & \multicolumn{1}{c|}{---} & 322.922 (1006$\times$) \\ \hline
Rollup of 100 Pvt Tx & 2\textsuperscript{26} & \multicolumn{1}{c|}{640987} & \multicolumn{1}{c|}{---} & 645.029 (994$\times$) \\ \hline
\end{tabular}%
}
\end{table}

\subsubsection{Performance Comparison}
\label{sec:perf_comp}
\autoref{tab:iso_zkspeed_comp} compares a 300 mm$^2$ zkPHIRE design using the same modular multipliers as zkSpeed+ (366 mm$^2$ with MLE Updates pipelined into SumCheck) and without ZeroCheck masking. This is to provide as fair a comparison as possible (zkSpeed+ is roughly $10\%$ faster than zkSpeed). Speedups are reported relative to CPU. zkPHIRE is about $10\%$ slower than zkSpeed+, while offering flexibility to support arbitrary, custom gates. \autoref{tab:iso_cpu_comp} shows zkPHIRE's speedups at iso-CPU area (294 mm$^2$) with fixed primes and ZeroCheck masking, achieving $1486\times$ geomean speedup and generating proofs for workloads upwards of $2^{30}$ nominal gates. Lastly,  \autoref{tab:iso_app_comp} compares the same design (with ZeroCheck masking and fixed primes) with zkSpeed+ at iso-application, achieving geomean $11.87\times$ speedup.

\begin{table}[]
\caption{Runtime and speedup compared to CPU for Jellyfish gates. Vanilla gate counts are shown to show original problem size. }
\label{tab:iso_cpu_comp}
\resizebox{\columnwidth}{!}{%
\begin{tabular}{|c|cc|cc|}
\hline
\multirow{2}{*}{Workload} & \multicolumn{2}{c|}{Gates} & \multicolumn{2}{c|}{Runtime (ms)} \\ \cline{2-5} 
 & \multicolumn{1}{c|}{Vanilla} & Jellyfish & \multicolumn{1}{c|}{CPU} & zkPHIRE \\ \hline
ZCash & \multicolumn{1}{c|}{2\textsuperscript{17}} & 2\textsuperscript{15} & \multicolumn{1}{c|}{701} & 0.750 (934$\times$) \\ \hline
Zexe Recursive Ckt & \multicolumn{1}{c|}{2\textsuperscript{22}} & 2\textsuperscript{17} & \multicolumn{1}{c|}{1951} & 1.440 (1354$\times$) \\ \hline
Rollup of 10 Pvt Tx & \multicolumn{1}{c|}{2\textsuperscript{23}} & 2\textsuperscript{18} & \multicolumn{1}{c|}{3339} & 2.269 (1471$\times$) \\ \hline
Rollup of 25 Pvt Tx & \multicolumn{1}{c|}{2\textsuperscript{24}} & 2\textsuperscript{19} & \multicolumn{1}{c|}{6161} & 3.874 (1590$\times$) \\ \hline
2\textsuperscript{12} Rescue Hashes & \multicolumn{1}{c|}{2\textsuperscript{21}} & 2\textsuperscript{20} & \multicolumn{1}{c|}{11532} & 7.114 (1621$\times$) \\ \hline
Rollup of 50 Pvt Tx & \multicolumn{1}{c|}{2\textsuperscript{25}} & 2\textsuperscript{20} & \multicolumn{1}{c|}{11533} & 7.114 (1621$\times$) \\ \hline
Rollup of 100 Pvt Tx & \multicolumn{1}{c|}{2\textsuperscript{26}} & 2\textsuperscript{21} & \multicolumn{1}{c|}{24071} & 13.614 (1474$\times$) \\ \hline
Rollup of 1600 Pvt Tx & \multicolumn{1}{c|}{2\textsuperscript{30}} & 2\textsuperscript{25} & \multicolumn{1}{c|}{355406} & 207.673 (1711$\times$) \\ \hline
zkEVM & \multicolumn{1}{c|}{---} & 2\textsuperscript{27} & \multicolumn{1}{c|}{25 min} & 828.948 (1809$\times$) \\ \hline
\end{tabular}%
}
\end{table}

\begin{table}[]
\caption{Iso-application comparison of zkSpeed+ (Vanilla)\\and zkPHIRE (Jellyfish).}
\label{tab:iso_app_comp}
\resizebox{\columnwidth}{!}{%
\begin{tabular}{|c|cc|cc|}
\hline
\multirow{2}{*}{Workload} & \multicolumn{2}{c|}{Gates} & \multicolumn{2}{c|}{Runtime (ms)} \\ \cline{2-5} 
 & \multicolumn{1}{c|}{Vanilla} & Jellyfish & \multicolumn{1}{c|}{zkSpeed+} & zkPHIRE \\ \hline
ZCash & \multicolumn{1}{c|}{2\textsuperscript{17}} & 2\textsuperscript{15} & \multicolumn{1}{c|}{1.825} & 0.750 (2.43$\times$) \\ \hline
2\textsuperscript{12} Rescue Hashes & \multicolumn{1}{c|}{2\textsuperscript{21}} & 2\textsuperscript{20} & \multicolumn{1}{c|}{19.631} & 7.114 (2.75$\times$) \\ \hline
Zexe Recursive Circuit & \multicolumn{1}{c|}{2\textsuperscript{22}} & 2\textsuperscript{17} & \multicolumn{1}{c|}{38.535} & 1.440 (26.76$\times$) \\ \hline
Rollup of 10 Pvt Tx & \multicolumn{1}{c|}{2\textsuperscript{23}} & 2\textsuperscript{18} & \multicolumn{1}{c|}{76.356} & 2.269 (33.65$\times$) \\ \hline
Rollup of 25 Pvt Tx & \multicolumn{1}{c|}{2\textsuperscript{24}} & 2\textsuperscript{19} & \multicolumn{1}{c|}{151.973} & 3.874 (39.23$\times$) \\ \hline
\end{tabular}
}
\end{table}

\section{Related Work}
\label{sec:Related}

\begin{table}[t]
\centering
\caption{Comparison of zkPHIRE with Prior ZKP Accelerators.\\N = NTT, S = SumCheck, M = MSM. Areas are in 7nm.}
\label{tab:comp_table}
\rowcolors{2}{rowgray}{white}
\resizebox{\columnwidth}{!}{%
\begin{tabular}{>{\bfseries}l|c c c c}
\hline
\textbf{Metric} & \textbf{NoCap} & \textbf{SZKP+} & \textbf{zkSpeed+} & \textbf{zkPHIRE} \\
\hline
Workload & Scaled-up AES & \begin{tabular}[c]{@{}c@{}}Rollup 25\\Pvt Tx\end{tabular} & \begin{tabular}[c]{@{}c@{}}Rollup 25\\Pvt Tx\end{tabular} & \begin{tabular}[c]{@{}c@{}}Rollup 25\\Pvt Tx\end{tabular}
\\
Protocol & Spartan+Orion & Groth16 & HyperPlonk & HyperPlonk \\
Main Kernels & N \& S & N \& M & S \& M & S \& M
\\
Gates & $2^{24}$ & $2^{24}$ & $2^{24}$ & $2^{19}$
\\
Encoding & R1CS & R1CS & \begin{tabular}[c]{@{}c@{}}Plonk\\ (Vanilla)\end{tabular} & \begin{tabular}[c]{@{}c@{}}Plonk\\ (Jellyfish)\end{tabular} \\
Proof Size & 8.1 MB & 0.18 KB & 5.09 KB & 4.41 KB \\
Setup & none & circuit-specific & universal & universal \\
Prime & fixed & arbitrary & arbitrary & fixed \\
Bitwidth & 64 & 255/381 & 255/381 & 255/381 \\
SW Prover (s) & 94.2 & 51.18 & 145.5 & 6.161 \\
HW Prover (ms) & 151.3 & 28.43 & 151.973 & 3.874 \\
SW Verifier (ms) & 134 & 4.2 & 26 & 19 \\
Chip Area (mm\textsuperscript{2}) & 38.73 & 353.2 & 366.46 & 294.32 \\
\# Modmuls & 2432 & 1720 & 1206 & 2267 \\
Modmul (mm\textsuperscript{2}) & 0.002 & 0.133 / 0.314 & 0.133 / 0.314 & 0.073/0.162 \\
Power (W) & 62 & $>$220 & 171 & 202 \\
\hline
\end{tabular}
}
\end{table}

Prior work on cryptographic acceleration has largely focused on Multi-Party Computation and Homomorphic Encryption \cite{kim2022_BTS, sharp, rpu, haac, karthik1, karthik2, ciflow, mo2023towards, ark, jha2024deepreshape, osiris, anaheim}.
Research on ZKP acceleration is relatively newer, and has focused primarily on MSMs and NTTs~\cite{priormsm, distmsm, cuzk, rezk, legozk, unizk, pipezk, gzkp, gypso, MSMAC, tches_ntt_msm, zhang2025code, zhang2025micro, elastic_msm, sam, myosotis, graz, prophet}.
The community has broadened its focus to include other ZKP kernels such as the SumCheck protocol and Merkle tree-based commitments.
Prior work has explored accelerating these kernels on GPUs~\cite{batchzk} and ASICs~\cite{nocap, zkspeed2025}.
We compare zkPHIRE with three ASICs that accelerate proof-generation end-to-end: SZKP, NoCap, and zkSpeed.

\textbf{Protocol Acceleration}:
\textbf{SZKP} is a Groth16~\cite{groth16} accelerator that speeds up all MSMs, including sparse G2 MSMs, achieving a geomean speedup of 493× over CPU. While a direct comparison is not apples-to-apples due to differing proof sizes and trusted setup assumptions, Groth16 and HyperPlonk target similar application domains requiring small, efficiently verifiable proofs. Unlike Groth16, HyperPlonk supports a universal setup and avoids an expensive trusted setup ceremony~\cite{ceremony}. Following zkSpeed, we report results for a scaled SZKP+ configuration that adopts more efficient MSMs and the BLS12-381 curve.
\textbf{NoCap} accelerates Spartan+Orion proofs using a vector architecture. It relies on Merkle tree-based commitments, resulting in larger proofs and higher verifier costs, and is better suited for settings where proof succinctness is less critical or number of verifiers is small.
\textbf{zkSpeed} is, to our knowledge, the only prior accelerator targeting HyperPlonk end-to-end, achieving a geometric mean speedup of 801× over CPU. For our comparison here, we use zkSpeed+ as we have done earlier. 

\autoref{tab:comp_table} summarizes the comparison across the four accelerators and reports iso-workload performance numbers.
We use numbers for a scaled-up AES workload from NoCap's evaluation as it has a similar number of gates to the Rollup 25 workload. 
We use the same area scaling factors used by zkSpeed to scale our area and NoCap's area to 7nm.
In our experiments, we assume a 1:1 mapping between R1CS constraints and Vanilla constraints. If Vanilla circuits require more constraints, zkPHIRE’s Jellyfish gains relative to Vanilla gates would be larger, scaling roughly linearly at iso-workload.
zkPHIRE's proving time is 39$\times$, 7$\times$, and 39$\times$ faster than NoCap, SZKP+, and zkSpeed+, respectively.
Jellyfish's advantages are application-dependent: not all workloads map efficiently to Jellyfish gates. In such cases, different custom gates (which zkPHIRE can support) may yield faster proving times. 
Finally, while SZKP+ still yields the smallest proofs, it suffers from a major usability drawback due to the costly circuit-specific trusted setup.

\textbf{Limitations of prior work}:
Prior work \cite{nocap} accelerates SumCheck using vector processors, but vector-style reductions introduce significant overhead: reductions over length-$M$ vectors require $\log_2(M)$  serialized folding steps, repeated register-file accesses, and additional reductions for higher-degree extensions. zkPHIRE avoids these bottlenecks by using fused, tree-structured reduction pipelines with minimal data movement.
While prior work \cite{nocap, legozk} can scale to workloads of $2^{30}$ constraints, their reliance on NTT-based matrix encodings introduces substantial latency overhead, e.g., off-chip transposes and recursive decompositions.
In contrast, zkPHIRE has straightforward memory accesses, and given the streaming architecture, can scale well beyond $2^{30}$ constraints, as long as the working set fits in off-chip memory. For ZKML or other problems of this scale, zkPHIRE would need better data orchestration schemes (e.g. fetching from the cloud). We leave this for future work.

\section{Conclusion}
This paper presents zkPHIRE, the first accelerator to support programmable SumCheck for succinct proof generation at very large scales, enabling efficient ZKPs over custom, high-degree gates.
On HyperPlonk baselines, zkPHIRE achieves iso-application geomean speedups of 
$11.87\times$ over prior ASICs and $1486\times$ over CPU.
Its design abstracts SumCheck into common primitives—updates, extensions, and products—and exploits parallelism within tables, across tables, and across evaluation products to scale across diverse polynomial structures.
By streaming polynomial tables through compact scratchpads and coupling compute units via a Multifunction Forest, zkPHIRE improves resource sharing and reduces SRAM and datapath overheads. Finally, a systematic design-space exploration balances parallelism, area efficiency, and performance to support a wide range of polynomials and custom gate constructions.

\section*{Acknowledgements}
This work was supported in part by DARPA’s Hybrid Electromagnetic Side-Channel and Interactive-Proof Methods to Detect and Amend LogicaL Rifts (HEIMDALLR) program (Grant HR0011-25-C-0300), NSF CAREER Award \#2340137, NSF CIRC GRAND \#2450539, an NSF RINGS Award, and a gift award from Google.
% \end{acks}
% This document is an updated version of HPCA 2022 and 2023, which, in
% turn, has been derived from two previous conferences, in particular
% HPCA 2021 and MICRO 2021, which, in turn, are derived from past MICRO,
% HPCA, ISCA, and ASPLOS conferences.

%%%%%%% -- PAPER CONTENT ENDS -- %%%%%%%%

%%%%%%%%% -- BIB STYLE AND FILE -- %%%%%%%%
\bibliographystyle{IEEEtranS}
\bibliography{refs}
%%%%%%%%%%%%%%%%%%%%%%%%%%%%%%%%%%%%

\end{document}